\documentclass[prb,twocolumn,showpacs,preprintnumbers,amsmath,amssymb]{revtex4}
\usepackage{graphicx}% Include figure files
\usepackage{dcolumn}% Align table columns on decimal point
\usepackage{bm}% bold math

\begin{document}   

\title{Manganites at Quarter Filling: Role of Jahn-Teller Interactions}

\author{Jan Ba{\l}a}
\affiliation{Max-Planck-Institut f\"ur Festk\"orperforschung,
             Heisenbergstrasse 1, D-70569 Stuttgart, Germany} 
\affiliation{Marian Smoluchowski Institute of Physics, Jagellonian University,
             Reymonta 4, PL-30059 Krak\'ow, Poland}
\author{Peter Horsch}
\author{Frank Mack}
\affiliation{Max-Planck-Institut f\"ur Festk\"orperforschung,
             Heisenbergstrasse 1, D-70569 Stuttgart, Germany}

\date{\today}

\begin{abstract}
We have analyzed different correlation functions in a realistic spin-orbital
model for half-doped manganites. Using a finite-temperature 
diagonalization technique 
the CE phase was found in the charge-ordered phase in the case of small 
antiferromagnetic interactions between $t_{2g}$ electrons. 
It is shown that a key ingredient responsible for stabilization of 
the CE-type spin and orbital-ordered state is the cooperative Jahn-Teller 
(JT) interaction between next-nearest Mn$^{+3}$ neighbors 
mediated by the breathing mode
distortion of  Mn$^{+4}$ octahedra and displacements of Mn$^{+4}$ ions.  
The topological phase factor in the Mn-Mn hopping leading to gap
formation in one-dimensional models for the CE phase as well as the
nearest neighbor JT coupling  are not able to produce the zigzag chains
typical for the CE phase in our model. 
\end{abstract}

\pacs{75.30.Kz, 71.70.Ej, 75.30.Et, 75.10.-b}

\maketitle

\section{\label{sec:intro}Introduction}

Half-doped perovskite manganites Re$_{1/2}$A$_{1/2}$MnO$_3$ 
(Re-rare earth, A-alkaline earth) exhibit very specific properties 
quite different from manganites with other $e_g$ electron 
concentrations.\cite{Ima98,Tok00} Generally, 
for large size Re and A cations 
or when external pressure is applied\cite{Mor97} these compounds are in 
the A-type antiferromagnetic (AF), metallic state at low 
temperatures showing no or only weak sign of charge ordering, and
the occupied $e_g$ electron states are predominantly $x^2-y^2$ like. 
Recently it has been reported that 
the A-phase may  develop an intrinsic charge-stripe
modulation, which controls the transport properties.\cite{Kajimoto02}
With decreasing size of the cations, which implies a decrease of the
bandwidth, charge ordering 
is observed which at lower temperatures is accompanied by the formation of 
peculiar ferromagnetic (FM) zigzag chains\cite{Ste96,Mur98,Nak99} 
that are staggered antiferromagnetically. 
This CE spin order, a notation introduced in the pioneering work
of  Wollan and Koehler\cite{Wol55} and  Goodenough\cite{Goo55}, is 
accompanied by a checkerboard charge order (CO) and directed occupied 
orbitals on the Mn$^{3+}$ sites amplifying double exchange (DE) along one
lattice diagonal. 
The CE-structure is a particularly fascinating manifestation of 
the control of magnetic order due to the orbital 
degree of freedom\cite{Kil99,Kug73} in combination with charge ordering.
Although a number of approaches have been followed to explain this structure,
the mechanism for its stability 
is still not clear and hotly debated with emphasis put on 
the role of the DE,\cite{Sol99,BKK99,Sol01}
the Coulomb interaction,\cite{Mut99,Jac00,Shu01,Pop02} or the coupling to the 
lattice degrees of freedom.\cite{Miz97,Yun00,Hot00,HFD01,Gu02,Wei02,Wil03}

The CE phase has been observed experimentally both in cubic 
(Nd,Pr)$_{1/2}$(Sr,Ca)$_{1/2}$MnO$_3$\cite{Tom95,Kaw97,Nak99,Tok00,Zim01}
and in layered LaSr$_2$Mn$_2$O$_7$\cite{Kub99} and 
La$_{1/2}$Sr$_{3/2}$MnO$_4$\cite{Ste96,Mur98} manganites. In some cases 
it can coexist with the A-type AF spin ordering.\cite{Kub99,Arg00,Ish00} 
Regarding CO in the CE phase, some experiments are interpreted in terms
of almost perfect CO \cite{Mur98,Che99} while others\cite{Lar01}  
are considered to be consistent with a smooth
charge density wave (CDW) of the $e_g$ electron density. Although, it is  
plausible that the complex spin and orbital 
arrangement in the charge-ordered manganites can
result from the competition between DE\cite{Zen51} and 
AF superexchange interactions ($\sim J_\textrm{AF}$), 
the role played by the CO and Jahn-Teller (JT)
interactions\cite{Yun00,Hot00,HFD01,HMD00} in the system is still unclear. 
Finally in systems like e.g. Pr$_{1/2}$Ca$_{1/2}$MnO$_3$ CO and CE-type
orbital correlations are found to develop well above the Neel 
temperature\cite{Kaj01,zim99} or even coexist with FM spin 
state,\cite{Geck02} which indicates that these orbital correlations 
are not primarily driven by magnetic interactions.

This work focuses on a microscopic understanding of the stability of
the CE-phase which appears to be the ground state of narrow band
manganites at and near quarter-filling.
We argue here that the key interaction, apart from the nearest-neighbor
Coulomb repulsion $V$ which supports the charge ordering,\cite{Mor99} 
is the next-nearest neighbor JT interaction. This interaction
between orbitals at next-nearest neighbor Mn$^{3+}$ ions emerges 
from the JT-distortions of the Mn$^{3+}$ octahedra which is amplified
by the breathing distortion of the intermediate Mn$^{4+}$ octahedra. 
We show that
this interaction generates a narrow regime in the $V$-$J_\textrm{AF}$
phase diagram where the CE-phase (and sometimes the C-phase) is stable. 
Otherwise  one encounters a homogeneous FM phase 
at small  $J_\textrm{AF}$ fully controlled by DE
and a conventional nearest-neighbor AF phase at larger  $J_\textrm{AF}$. 

An important feature of  further neighbor JT 
interactions are the displacements of 
the Mn$^{4+}$ ions. We are only aware of work by 
Radaelli \textit{et al.}\cite{Rad97} where  
detailed lattice coordinates for the CE-phase in La$_{1/2}$Ca$_{1/2}$MnO$_3$
are reported. In particular they observed that the 4 planar O(2) 
move towards the 
Mn$^{4+}$ ion and the  Mn$^{4+}$ ions are found to be displaced from their 
regular perovskite positions, while the Mn$^{3+}$ ions are not shifted. 
This observation confirms the importance  of the breathing 
distortion and also 
shows the mechanism of the further neighbor JT-distortion in the CE-phase.
As a consequence the appearance of new Bragg reflections in the CE-phase 
would be associated not primarily with quadrupolar electronic orbital order
but rather with the displacements of ions associated with the 
CE orbital order.\cite{Rad97}

An alternative long-ranged JT based interaction was recently
proposed by Khomskii and Kugel\cite{Kho01} and by Calderon, Millis and
Ahn.\cite{Cal03}
They employ elasticity theory to derive the orbital interaction between 
JT-distorted Mn$^{3+}$O$_6$ octahedra for
the perfectly charge ordered case. We have also investigated 
the consequences of this interaction. Our numerical simulations 
confirm the estimate presented in Ref.~\onlinecite{Kho01}
that for perfect charge ordering the AF CE phase is stable.
Yet in this model this requires large nearest neighbor 
Coulomb repulsion $V$, 
while for moderate values of $V$ the CE phase gets unstable.

The finite temperature diagonalization\cite{Jac00,Hor99,Mac99} 
used in our study
allows to include the full interplay of spin and orbital states,
and to monitor the onset of different orders as function of temperature. 
This gives us the ability for an unbiased investigation of the formation 
of different
spin-orbital orderings emerging from all multi-electron configurations
in the cluster. 
In this paper we concentrate on  
the stability and temperature dependence of spin, orbital, and charge 
ordered states in the two-dimensional (2D) realistic model for half-doped
manganites. In our microscopic spin-orbital model CO develops as a result 
of nearest-neighbor Coulomb repulsion $V$ and as a consequence of further
neighbor JT interactions, while 
the spin-orbital order results from the competition between 
DE (i.e. kinetic energy), AF superexchange and 
further neighbor JT interactions. 
As shown in Sec.~\ref{sec:nnnjt} the latter interaction can  lead 
to a CO state at low temperatures even in the absence of inter-site 
Coulomb repulsion. 
Furthermore we argue that  a topological sign in the 
hopping amplitudes of $e_g$ 
orbitals\cite{BKK99} invoked in one-dimensional (1D) models for the CE phase 
is not sufficient to explain 
the formation of CE spin-orbital structure in the 2D model studied by us. 

Further motivations to put  effort in a better understanding
of the interplay of spin, charge, orbitals and lattice in the formation
of the CE-phase are the following: It has been argued that the intrinsic
mechanism that leads to colossal magnetoresistance are tendencies towards
CE-type charge and orbital ordering in the vicinity of 
$x=0.5$.\cite{Ueh99}
Furthermore a CE-insulator to FM-metal transition has been observed 
in relatively small magnetic field, 
which is accompanied by a major change of the
optical conductivity on a surprisingly large energy scale 
($\sim 1$ eV).\cite{Oki99,Jun00} Such dramatic changes in relatively weak 
magnetic fields are a clear manifestation of the subtle interplay 
between spin, orbital and CO.
The present study provides a basis for further investigations of the
temperature dependence of e.g. the optical conductivity and the
study of spin and orbital excitations in the CE-phase, as well as
the effect of doping into the quarter filled state. 

The paper is organized as follows. In Sec.~\ref{sec:model} we present 
the model 
Hamiltonian, and describe in Sec.~\ref{sec:lanczos}  the  
finite-temperature Lanczos method. 
Sec.~\ref{sec:numerics} contains numerical 
results for different two-site 
correlation functions characterizing the charge, orbital and spin structure.
These correlation functions are
evaluated as functions of temperature and collected in a form 
of semi-quantitative phase diagrams in three different regimes: (i) including
nearest-neighbor effective orbital-orbital (OO) interactions 
(Sec.~\ref{sec:nnjt}); (ii) assuming the 
elastic form of OO coupling at different distances (Sec.~\ref{sec:elastic}); 
(iii) considering the local form of next- and nearest-neighbor OO coupling
(Sec.~\ref{sec:nnnjt}), while in Sec.~\ref{sec:stacking} the effect of 
charge stacking is discussed. The paper is summarized in 
Sec.~\ref{sec:summary}. In the Appendix the stability of the CE versus 
the C phase is
analysed for a 1D band model with lifted orbital degeneracy.

%%%%%%%%%%%%%%%%%%%%%%%%%%%%%%%%%%%%%%%%%%%%%%%%%%%%%%%%%%%%%%%%%%%%%%%%%
\section{\label{sec:model}Model Hamiltonian and numerical procedure}
%%%%%%%%%%%%%%%%%%%%%%%%%%%%%%%%%%%%%%%%%%%%%%%%%%%%%%%%%%%%%%%%%%%%%%%%%

\subsection{\label{sec:eff_model}Effective spin-orbital model}

We consider the generic FM Kondo lattice model for 
manganites\cite{Hor99,Mac99} with
the Mn $e_g$ electrons coupled to the core $t_{2g}$ spins via the Hund
coupling $J_H$. The model is augmented  by
inter-site Coulomb repulsion $V$ and cooperative JT interactions 
and treated in the limit
of infinite on-site Coulomb repulsion between two $e_g$ electrons on 
the same site ($U\to \infty$):
\begin{equation}
H = H_{band} + H_{Kondo} + H_\textrm{AF} + H_{V} + H_{OO}.
\label{eq:H_total}
\end{equation}
The first term describes the motion of the $e_g$ electrons,
\begin{equation}
H_{band} = -\sum_{\langle {\bf ij}\rangle \xi\zeta,\sigma}
\left(t^{\xi\zeta}_{\bf ij}\tilde{d}^{\dagger}_{{\bf i}\xi\sigma}
\tilde{d}_{{\bf j}\zeta\sigma}
+ \textrm{H.c.}\right),
\end{equation}
with a constraint that allows only for Mn$^{4+}$ and Mn$^{3+}$ configurations, 
$\tilde{d}^{\dagger}_{{\bf i}\xi\sigma}=d^{\dagger}_{{\bf i}\xi\sigma}
(1 - n_{{\bf i}\xi\bar{\sigma}})
\prod_{\sigma'}(1-n_{{\bf i}\bar{\xi}\sigma'})$,
where all other alternative $e_g$ states are projected out. Here and in the 
following the index $\bar{\xi}$ ($\bar{\sigma}$) denotes the $e_g$ orbital 
(spin) orthogonal to the $\xi$ ($\sigma$) one, respectively. 
The hopping matrix elements depend on the basis chosen. For the basis
$\{|x\rangle,|z\rangle\}$ with $|x\rangle\sim x^2-y^2$ 
and $|z\rangle\sim 3z^2-r^2$ they are given by,
\begin{equation}
\label{eq:t}
\left[t^{\xi\zeta}_{{\bf ij}||a(b)}\right]=\frac{t}{4}\left(
\begin{array}{cc}
3 & \mp\sqrt{3}\\
\mp\sqrt{3} & 1
\end{array}\right) ,
\end{equation}
where, $+$($-$) refer to the $a$($b$) direction, respectively. The second 
term of Hamiltonian (\ref{eq:H_total}) stands for the Hund interaction 
between $e_g$ electrons spin and the $S=3/2$ $t_{2g}$ 
core spin ${\bf S}_{\bf i}$,
\begin{equation}
\label{eq:H_kondo}
H_{Kondo} = -J_H\sum_{{\bf i} \xi\sigma\sigma'}
{\bf S}_{\bf i}\cdot\tilde{d}^{\dagger}_{{\bf i}\xi\sigma}
\vec{\sigma}_{\sigma\sigma'}  
\tilde{d}_{{\bf i}\xi\sigma'} ,
\end{equation}
resulting in the parallel alignment of spins between $t_{2g}$ and $e_g$
electrons at each site in the limit $J_H\gg t$ which 
corresponds to the realistic situation in manganites. The
$H_{band} + H_{Kondo}$ part of the total Hamiltonian alone
represents the DE mechanism and yields
a FM ground state with $|x\rangle$ orbitals being occupied which is
favored by the kinetic energy in 2D model. Such a fully spin polarized case 
at finite $U$ leads to \textit{the orbital $t$-$J$ model} as considered 
in Refs.~\onlinecite{Hor99,Mac99}. 
Here, we are interested in the more
complex case where both spin and orbital degrees of freedom play an active 
role. Therefore, an AF ($J_\textrm{AF}>0$) coupling between 
nearest-neighbor $t_{2g}$ spins is also incorporated, 
\begin{equation}
\label{eq:H_AF}
H_\textrm{AF} = J_\textrm{AF}\sum_{\langle{\bf ij}\rangle}
{\bf S}_{\bf i}\cdot{\bf S}_{\bf j} .
\end{equation}
In our numerical studies we assume $S=1/2$ for the core spins which 
decreases the resulting 
Hilbert space considerably. We have tested that this assumption gives similar
results as in the $S=3/2$ case when $J_H$ is renormalized\cite{Mack99}
by the factor $3$.
Moreover, to promote the charge ordering as observed experimentally in 
half-doped manganites we include the Coulomb repulsion ($V>0$) between $e_g$ 
electrons on neighboring sites,\cite{Mack99}
\begin{equation}
\label{eq:H_V}
H_V = V\sum_{\langle{\bf ij}\rangle\xi\zeta\sigma\sigma'}
n_{{\bf i}\xi\sigma}n_{{\bf j}\zeta\sigma'} ,
\end{equation}
with $n_{{\bf i}\xi\sigma}=\tilde{d}^{\dagger}_{{\bf i}\xi\sigma}
\tilde{d}_{{\bf i}\xi\sigma}$. This term reduces the probability of
occupancy of neighboring sites by $e_g$ electrons.

The form of the inter-site OO interactions included in
the $H_\textrm{OO}$ term of the Hamiltonian will be specified in 
the following section.

\subsection{\label{sec:lanczos}Finite-temperature Lanczos method}

To investigate the possible spin and orbital orderings in our spin-orbital  
model 
(\ref{eq:H_total}) we have calculated the temperature dependence of different
two-site correlation functions given by the operator expectation value:
\begin{equation}
\langle A_{\bf R}\rangle 
= \frac {1}{Z}\sum_{n=1}^{N_{st}}\langle n|\exp(-\beta H)A_{\bf R}|n\rangle ,
\label{eq:A_R_exact}
\end{equation}
of two-site operators, $A_{\bf R}=1/N\sum_{\bf i}B_{\bf i}B_{{\bf i}+{\bf R}}$
with $B_{\bf i}=n_{\bf i}$, $T^z_{\bf i}$, $S^z_{\bf i}$, or
$S^z_{\bf i}n_{\bf i}$ with $n_{\bf i}=\sum_{\xi,\sigma}
n_{{\bf i}\xi\sigma}$.\cite{Mack99} 
Here, $\beta=1/k_BT$, $N_{st}$ is the number of 
basis states $\{|n\rangle\}$ that span $H$, $N$ number of sites in the cluster
and $Z=\sum_{n=1}^{N_{st}}\langle n|\exp(-\beta H)|n\rangle$ is the partition 
function. 

The calculation of (\ref{eq:A_R_exact}) would, however, require a complete 
diagonalization
of a $N_{st}\times N_{st}$ matrix of the Hamiltonian. 
Therefore, we use for the 
calculation of $\langle A_{\bf R}\rangle$ a generalization of the exact 
diagonalization technique developed by 
Jakli\v{c} and Prelov\v{s}ek.\cite{Jak00}
In this approach the trace of the thermodynamic expectation value is performed 
by a Monte Carlo sampling leading to the following approximate expression:
\begin{equation}
\langle A_{\bf R}\rangle \approx \frac{N_{st}}{LZ}\sum_{l=1}^{L}\sum_{j=1}^{M}
\exp(-\beta E_j^l)\langle l|\psi^l_j\rangle\langle\psi_j^l|A_{\bf R}|l\rangle ,
\label{eq:A_R}
\end{equation}
where the first sum is performed over a restricted number $L$ of random states
$|l\rangle$ being initial states in the Lanczos
algorithm to generate a set $\{\psi^l_j\}$ of eigenfunctions of $H$ with
respective eigenvalues $\{E^l_j\}$. $M$ is the number of Lanczos functions
in the expansion of a given state $|l\rangle$. In the same manner 
the partition function is approximated by,
\begin{equation}
Z \approx \sum_{l=1}^{L}\sum_{j=1}^{M}
\exp(-\beta E_j^l)|\langle l|\psi^l_j\rangle|^2 .
\end{equation}
It has been shown\cite{Jak00} that the results get very accurate already for 
$L$, $M\ll N_{st}$. Even though we exploit the translational symmetry 
the calculations are restricted to small
clusters ($N=\sqrt{8}\times\sqrt{8}$ is used in this paper) as only 
$S^z_{tot}=\sum_{\bf i}^NS^z_{\bf i}$ subspaces can be treated separately,
while a similar symmetry in the orbital sector does not exist.
Although a $\sqrt{8}\times\sqrt{8}$ cluster is large enough to capture 
the CE-type correlations, boundary conditions can influence further
neighbor interactions. To assess such finite size effects quantitatively, a full
diagonalization on larger clusters would be necessary. In the case of the CO CE phase
this would involve at least a $4\times 4$ site cluster, which is far beyond 
numerical realization.

%%%%%%%%%%%%%%%%%%%%%%%%%%%%%%%%%%%%%%%%%%%%%%%%%%%%%%%%%%%%%%%%%%%%%%%%%%%%%
\section{\label{sec:numerics}Numerical results}
%%%%%%%%%%%%%%%%%%%%%%%%%%%%%%%%%%%%%%%%%%%%%%%%%%%%%%%%%%%%%%%%%%%%%%%%%%%%%

\subsection{\label{sec:parameters}Parameters and magnetic phases}

In the numerical studies we assume $S=1/2$ for the core spins and
the limit of strong Hund's coupling $J_H/t=15$. The numerical results
are almost identical for all $J_H\agt 5t$ in the whole range of 
temperatures considered by us. The Coulomb repulsion $V/t$ 
and the AF-exchange $J_\textrm{AF}/t$ are free parameters, and the data 
is summerized in form of a phase diagram in the $V-J_{AF}$ plane.
The hopping parameter $t$ is chosen as energy unit. 
We focus in this study on CO states and thus 
assume finite positive values for $V$ in all the calculations. In perovskite 
manganites the effective hopping $t$ strongly depends on the lattice 
parameters and distortions \textit{e.g.} increases with contraction of 
the Mn$-$O bond length and also depends on the Mn$-$O$-$Mn bond 
angle.\cite{Mor97} Thus, its value can be only roughly estimated as 
$t=0.2 - 0.7$eV.\cite{hop,Ish97,Fei99,Hor99} 

\begin{figure}
\includegraphics[width=6cm]{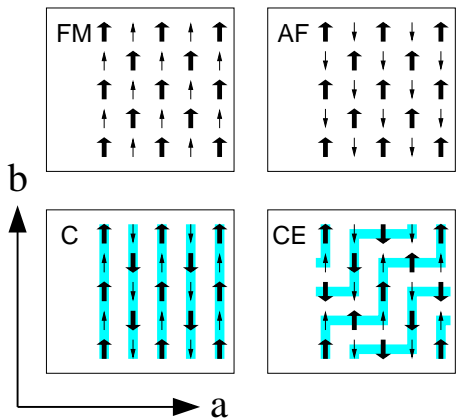}
\caption{\label{fig:spins}
Schematic view of different spin patterns found in the 2D model
in presence of checkerboard CO (half-doped case).
Thick (thin) arrows represent states with total 
spin $S=2$ ($S=3/2$) with (without) an $e_g$ electron
present at a given site, respectively. 
The shaded lines in C and CE phases indicate
the direction of DE carrier propagation. 
The preferred directional 
$e_g$ orbitals occupied at sites $S=2$ are parallel to the shaded paths.}
\end{figure}

A major difficulty when searching for spin structures like those indicated
in Fig.~\ref{fig:spins}, and the same applies to charge and orbital order, 
is that in a cluster 
with periodic boundary conditions translational  (and rotational) symmetries
are not broken. Therefore we have to investigate correlation functions and 
compare their features with the different spin configurations 
(presented in Fig.~\ref{fig:spins}).

%%%%%%%%%%%%%%%%%%%%%%%%%%%%%%%% FIG. 1 %%%%%%%%%%%%%%%%%%%%%%%%%%%%%%%%%%%%
%%%%%%%%%%%%%%%%%%%%%%%%%%%%%%%%%%%%%%%%%%%%%%%%%%%%%%%%%%%%%%%%%%%%%%%%%%%%

We shall consider three different 
forms for the orbital interaction term $H_\textrm{OO}$ and show 
that one has to go beyond the nearest-neighbor interactions to obtain the 
CE-type correlations. 
We begin the description of our calculations with a simplified model (without
$H_{\rm OO}$ term) but including the crystal-field splitting of $e_g$ orbitals
$\propto E_z$,
\begin{equation}
H_z = -E_z\sum_{\bf i}T^z_{\bf i} ,
\end{equation}
where $T^z_{\bf i}=\frac{1}{2}\sum_{\sigma} 
(n_{{\bf i}x\sigma}-n_{{\bf i}z\sigma})$ 
is the $z$ component of the pseudospin operator.
This  orbital splitting accounts for 
the elongation of octahedra along the
c-axis in layered compounds, which favors  $|z\rangle$ over  $|x\rangle$
orbital occupation, and also counterbalances the trend towards occupation
of  $|x\rangle$ orbitals, which is strongly favored by the kinetic energy
in the 2D-model. Although, the model without $H_{\rm OO}$ interactions
can lead to the stability of the CE-type orbital pattern,\cite{Mack99}
the magnetic structure was FM.
In a wide range of parameters $(E_z/t<0$, $V/t>0$, $J_\textrm{AF}/t>0)$ 
no stabilization of the AF CE spin phase 
was found, indicating that the inter-chain coupling is quite strong and 
cannot be neglected in realistic models for half-doped CO manganites. Such a 
conclusion agrees this recent density-functional calculations showing a 
considerable band dispersion normal to the chains.\cite{Pop02}
Thus, we argue that the \textit{topological effect} considered in 
1D model approximations, which gave a simple explanation of 
the existence of the CE spin-orbital phase at half-doping, is not likely to 
be the decisive mechanism in a more realistic 2D case investigated here.
Our simulations with purely electronic interactions do not provide
evidence for 1D FM chains that are coupled antiferromagnetically
to neighbor chains, which is the basic assumption that could justify
such 1D models.
Therefore, in the following sections we concentrate on the electron-lattice 
coupling\cite{Hot00,Yun00,HFD01,Mil98,BOS02} and present 
a mechanism stabilizing the CE spin/orbital state in manganites.

%%%%%%%%%%%%%%%%%%%%%%%%%%%%%%%% FIG. 2 %%%%%%%%%%%%%%%%%%%%%%%%%%%%%%%%%%%%
\begin{figure}
\includegraphics[width=3.5cm]{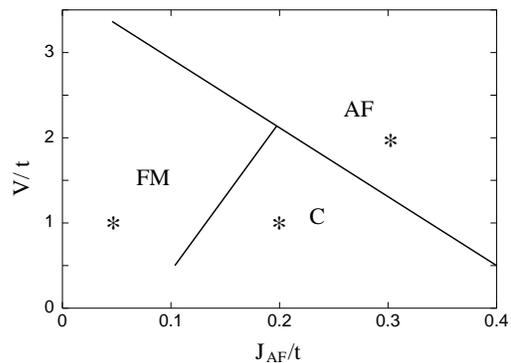}
\caption{\label{fig:OONN_phase}
Phase diagram indicating the character of the dominant 
spin correlations found at small temperatures $T<zJ$. 
The calculations include the nearest-neighbor OO effective  
interactions ($\kappa=0.2t$). '$*$' indicates representative points for which
the correlation functions are presented below.} 
\end{figure}
%%%%%%%%%%%%%%%%%%%%%%%%%%%%%%%%%%%%%%%%%%%%%%%%%%%%%%%%%%%%%%%%%%%%%%%%%%%%

\subsection{\label{sec:nnjt}The role of nearest-neighbor orbital interactions}

The first aim of our study of the spin-orbital model (\ref{eq:H_total}) in
half-doped manganites is to investigate the role of nearest-neighbor OO 
coupling played in CO state. 
There are two different mechanisms contributing to this interaction:
(i) the cooperative JT effect\cite{Fei99} and 
(ii) superexchange interactions.\cite{Hor99,Bri99} Neglecting
more complex spin-orbital terms\cite{Ish97,Fei99} both effects 
can be described by the following simple term derived before for undoped 
LaMnO$_3$ compound,\cite{Bri99}
\begin{equation}
\label{eq:H_OO}
H_\textrm{OO} = 2\kappa\sum_{\langle\bf ij\rangle}T_{\bf ij} ,
\end{equation}
where the two-site orbital operator, $T_{\bf ij}$, between nearest-neighbor 
Mn sites in the $\{|x\rangle,|z\rangle\}$ basis,
\begin{equation}
\label{eq:Tij}
T_{\bf ij} = T_{\bf i}^zT_{\bf j}^z + 3T_{\bf i}^xT_{\bf j}^x
\mp \sqrt{3}(T_{\bf i}^xT_{\bf j}^z + T_{\bf i}^zT_{\bf j}^x) ,
\end{equation}
is described in terms of pseudospin operators: 
$T^{+}_{\bf i}=\sum_{\sigma}\tilde{d}^{\dagger}_{{\bf i}x\sigma}
\tilde{d}_{{\bf i}z\sigma}$, and
$T^{-}_{\bf i}=\sum_{\sigma}\tilde{d}^{\dagger}_{{\bf i}z\sigma}
\tilde{d}_{{\bf i}x\sigma}$.
The prefactor of the mixed term $\propto\sqrt{3}$ is negative in the $a$
direction and positive in the $b$ direction. The nearest-neighbor OO term 
can lead to \textit{attraction} of electrons on 
neighboring sites with different orbital orientations but having the 
Coulomb repulsion (\ref{eq:H_V}) included in the model we 
avoid phase separation.

%%%%%%%%%%%%%%%%%%%%%%%%%%%%%%%% FIG. 3 %%%%%%%%%%%%%%%%%%%%%%%%%%%%%%%%%%%%
\begin{figure}
\includegraphics[width=10cm]{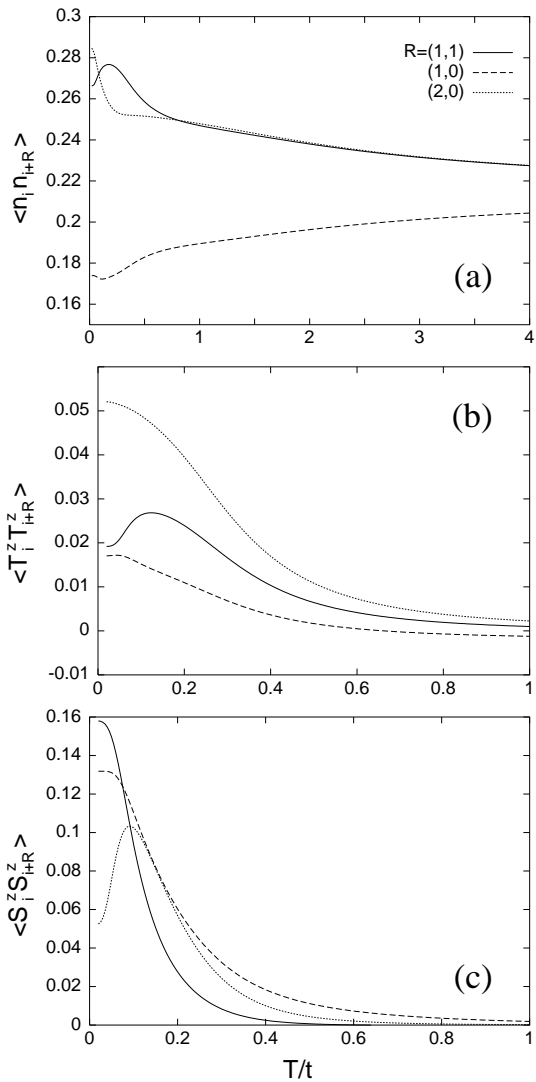}
\caption{\label{fig:corr_FM_NN}
Temperature dependence of the charge (a), orbital (b), and
spin (c) two-site correlation functions calculated for neighbors at
different distances ${\bf R}$ for the phase with FM spin correlations. 
Parameters: $V=t$, $\kappa=0.2t$, $J_\textrm{AF}=0.05t$, $J_H=15t$.}
\end{figure}
%%%%%%%%%%%%%%%%%%%%%%%%%%%%%%%%%%%%%%%%%%%%%%%%%%%%%%%%%%%%%%%%%%%%%%%%%%%%

%%%%%%%%%%%%%%%%%%%%%%%%%%%%%%%% FIG. 4 %%%%%%%%%%%%%%%%%%%%%%%%%%%%%%%%%%%%
\begin{figure}
\includegraphics[width=10cm]{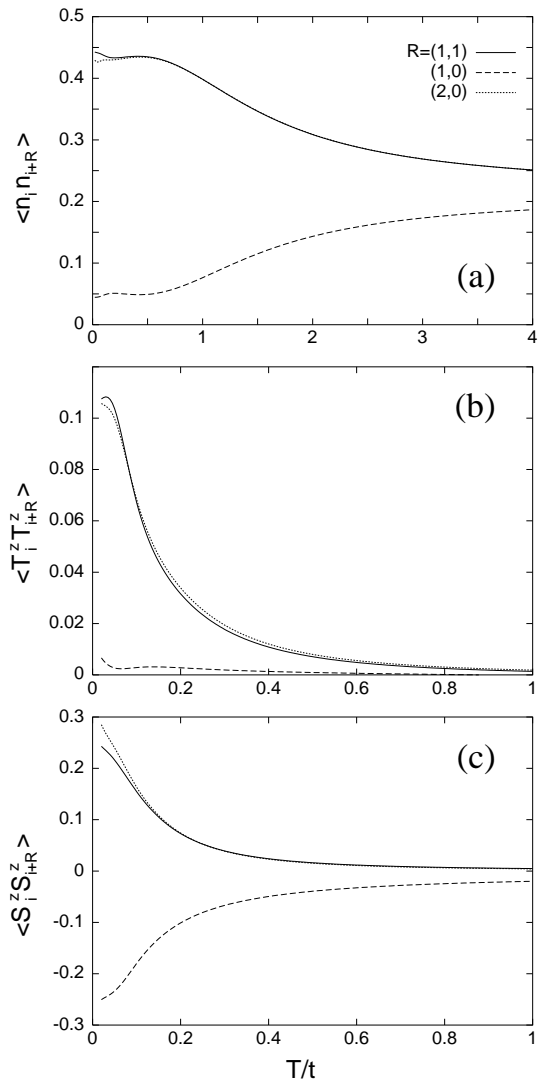}
\caption{\label{fig:corr_AF_NN}
Temperature dependence of the correlations as in 
Fig.~\protect{\ref{fig:corr_FM_NN}} but for the phase with AF spin 
correlations. Parameters: $V=2t$, $\kappa=0.2t$, $J_\textrm{AF}=0.3t$, 
$J_H=15t$.}
\end{figure}
%%%%%%%%%%%%%%%%%%%%%%%%%%%%%%%%%%%%%%%%%%%%%%%%%%%%%%%%%%%%%%%%%%%%%%%%%%%%

%%%%%%%%%%%%%%%%%%%%%%%%%%%%%%%% FIG. 5 %%%%%%%%%%%%%%%%%%%%%%%%%%%%%%%%%%%%
\begin{figure}
\includegraphics[width=10cm]{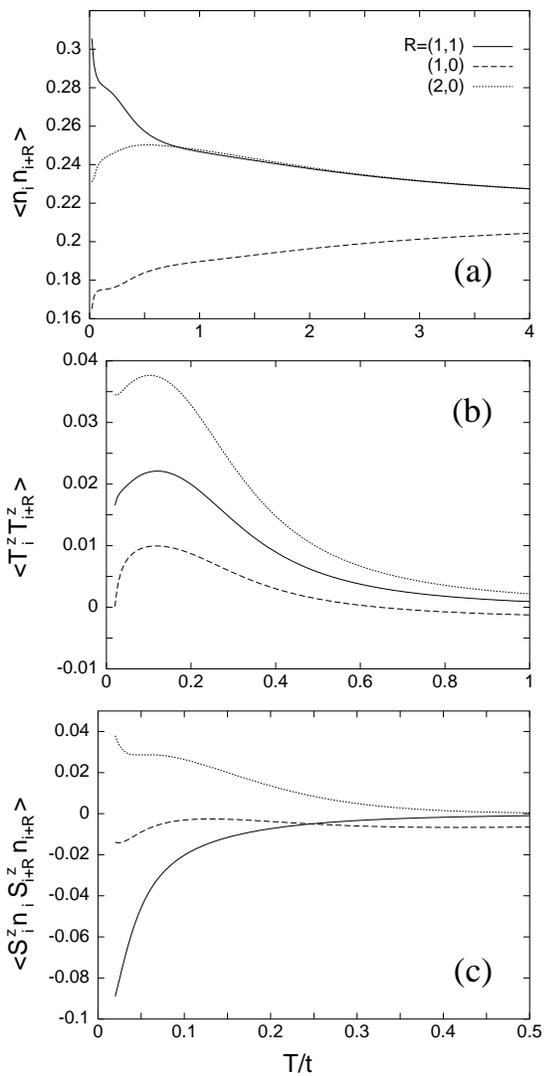}
\caption{\label{fig:corr_C_NN}
Temperature dependence of the charge (a), orbital (b), and
spin-charge (c) correlation functions calculated for the phase 
with C spin correlations. 
Parameters: $V=t$, $\kappa=0.2t$, $J_\textrm{AF}=0.2t$, $J_H=15t$.}
\end{figure}
%%%%%%%%%%%%%%%%%%%%%%%%%%%%%%%%%%%%%%%%%%%%%%%%%%%%%%%%%%%%%%%%%%%%%%%%%%%%

As shown in Fig.~\ref{fig:OONN_phase} for large values of the superexchange  
interaction between core spins ($zJ_\textrm{AF}\sim t$) 
an AF spin ground state is
realized while with decreasing $J_\textrm{AF}$ the kinetic energy of the $e_g$
electrons ($\sim t$) starts to play an active role via 
the DE mechanism\cite{Zen51} 
breaking some of the AF bonds which leads to 1D FM chains which form 
straight lines (C phase). Further decrease of 
$J_\textrm{AF}$ yields FM correlations in both $a$ and $b$ directions. 
As the kinetic energy is controlled by the inter-site Coulomb repulsion $V$
and suppressed in the limit of $V\gg t$ we find the AF region increasing in 
size with increasing $V/t$. 

We discuss now in more detail the temperature dependence of the different 
correlation functions.
Starting from the FM region (see Fig.~\ref{fig:corr_FM_NN}) we find
the evolution of weak CO for $T/t\alt 1$ with nearest-neighbor $(1,0)$ 
charge correlations smaller than the $(1,1)$ and $(2,0)$ next-neighbor 
ones which is consistent with the alternating 
order of Mn$^{+3}$ and Mn$^{+4}$ ions. The orbital correlations are 
positive between 
all neighbors indicating the preference of $x^2-y^2$ orbital occupancy due 
to the kinetic energy which is quite large in the FM state. Although all spin 
correlation functions are positive, they remain highly anisotropic
as $T\to 0$ as a result of the strong competition between the
DE mechanism and AF superexchange interaction between core spins.

Next, we consider the AF spin state in the case of large 
$J_\textrm{AF}$ 
and $V$  (see Fig.~\ref{fig:corr_AF_NN}). 
Increasing Coulomb repulsion ($V=2t$) we find more distinct
and almost isotropic alternation of the $e_g$ charge which now sets in 
already at $T/t\approx 2$. Here, the orbital correlations are  
correlated with the onset of the AF spin order at $T/t\approx 0.5$. 
For $T/t\to 0$ the spin-spin correlations are strong and represent an almost 
isotropic AF spin state.

For intermediate values of $J_\textrm{AF}$ and moderate Coulomb 
repulsion we find 
the phase with predominantly C spin correlations at $T/t\to 0$ (see Fig.
\ref{fig:OONN_phase}). As presented in Fig.~\ref{fig:corr_C_NN} 
for $J_\textrm{AF}/t=0.2$ and $V/t=1$ the charge 
correlations are similar to those obtained for the FM state shown in 
Fig.~\ref{fig:corr_FM_NN} (a) while the spin ordering changes dramatically. 
Moreover for $T/t\to 0$ the orbital correlations are weaker than those 
found in the FM state discussed above
indicating the orbital tendency towards in-plane directional-type order.

A convenient way to distinguish between phases with C or CE and other spin 
correlations is to evaluate a {\it combined spin-charge correlation function} 
being restricted to sites occupied by $e_g$ electrons,
$\langle S^z_{\bf i}n_{\bf i}S^z_{{\bf i}+{\bf R}}n_{{\bf i}+{\bf R}}\rangle$.
Analyzing the localized limit presented in Fig.~\ref{fig:spins} 
(or its by $\pi/2$ rotated version) one can easily see that, for the C phase 
one should find 
$\langle S^z_{\bf i}n_{\bf i}S^z_{{\bf i}+{\bf R}}n_{{\bf i}+{\bf R}}\rangle
>0$ for $|{\bf R}|=2$ and 
$\langle S^z_{\bf i}n_{\bf i}S^z_{{\bf i}+{\bf R}}n_{{\bf i}+{\bf R}}\rangle
<0$ for $|{\bf R}|=\sqrt{2}$ while for the CE phase only 
$\langle S^z_{\bf i}n_{\bf i}S^z_{{\bf i}+{\bf R}}n_{{\bf i}+{\bf R}}\rangle
<0$ for $|{\bf R}|=2$ with any other spin-charge correlations remaining small.
In the case presented in Fig.~\ref{fig:corr_C_NN} 
$\langle S^z_{\bf i}n_{\bf i}S^z_{{\bf i}+{\bf R}}n_{{\bf i}+{\bf R}}\rangle$
functions are
negative (positive) for nearest-neighbors along diagonal 
(next-nearest-neighbors along $a$/$b$ direction), respectively,  
in agreement with the arguments presented above. 
Furthermore, nearest-neighbor 
spin-charge correlations are small as a result of averaging over different 
cluster orientations (rotated by angles $\pm\pi/2$, $\pm\pi$).          

It is straightforward to see why the C-structure is favored by 
the nearest-neighbor JT 
interaction. The C-phase can be viewed as a realization of the alternating 
orbital structure of LaMnO$_3$ but with a modulated charge density. 
Hence this structure is compatible with  the nearest-neighbor JT 
interaction, whereas in the CE-structure this interaction is frustrated. 

\subsection{\label{sec:elastic}Elastic interactions}

\begin{figure}
\includegraphics[width=6cm]{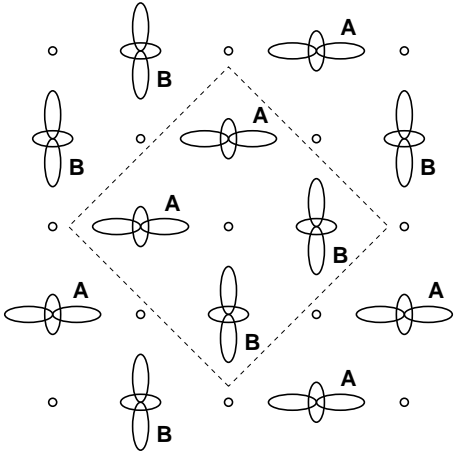}
\caption{\label{fig:ce_orb}
Schematic configuration of the charge-orbital ordered CE phase. The sites
{\bf A} and {\bf B} indicate two orbital Mn$^{3+}$ sublattices while
empty circles stand for the Mn$^{4+}$ ions. The dashed-line box indicates 
a typical orbitals orientation around a given Mn$^{4+}$ ion.}
\end{figure}

The role of local lattice distortions due to the JT effect, which creates an 
anisotropic strain field decaying as $\sim R^{-3}$ and thereby 
promotes long-range
orbital ordering, was considered by Khomskii (see Ref.~\onlinecite{Kho01}).
Following these arguments, we treat  the OO term as
an elastic interaction which is active also between $e_g$ electrons on 
next-nearest-neighbor sites along $a$ or $b$ direction,
\begin{equation}
H_{el} = 2\kappa\sum_{\langle\bf ij\rangle}T_{\bf ij} +
2\kappa'\sum_{\langle\langle{\bf ij}\rangle\rangle}T_{\bf ij} ,
\end{equation}
where $\langle\langle{\bf ij}\rangle\rangle$ refers to next nearest-neighbors 
along $a$ and $b$ direction with $\kappa'=\kappa/8$ which follows from the  
$\sim(|{\bf i}-{\bf j}|)^{-3}$ decay of elastic interactions.\cite{Kho01} 
The existence of 
longer-range OO interactions is justified by its cooperative lattice-driven
nature where an orbital flip on a single Mn site distorts not only the 
neighboring oxygens but also positions of other manganese ions around it. 
Here, the JT coupling 
preferring the tilting of the orbitals from the uniform orbital state 
leads to an alternating orbital order (see Fig.~\ref{fig:ce_orb})
and in this way compensates for 
the kinetic energy gain in uniform $|x\rangle$ orbital state favored in 
the 2D model. 
Here we neglect non-central elastic interactions between neighboring Mn sites 
in $(11)$ direction as their strength does not scale with $\kappa'$ 
(see Ref.~\onlinecite{Kho01}) and thus  would involve 
introduction of an additional parameter in the model. 

%%%%%%%%%%%%%%%%%%%%%%%%%%%%%%%% FIG. 6 %%%%%%%%%%%%%%%%%%%%%%%%%%%%%%%%%%%%
%%%%%%%%%%%%%%%%%%%%%%%%%%%%%%%%%%%%%%%%%%%%%%%%%%%%%%%%%%%%%%%%%%%%%%%%%%%%

%%%%%%%%%%%%%%%%%%%%%%%%%%%%%%%% FIG. 7 %%%%%%%%%%%%%%%%%%%%%%%%%%%%%%%%%%%%
\begin{figure}
\includegraphics[width=3.5cm]{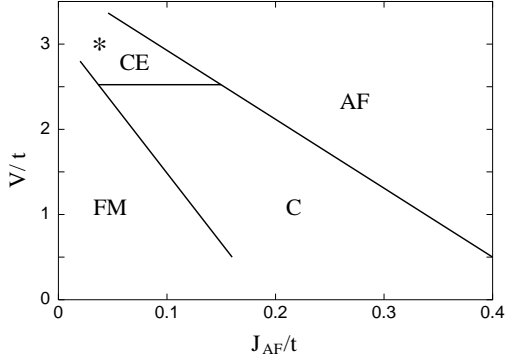}
\caption{\label{fig:elastic_phase}
Phase diagram as in Fig.~\protect{\ref{fig:OONN_phase}} including
the nearest- and next-neighbor elastic interactions ($\kappa=0.2t$, 
$\kappa'=\kappa/8$). '$*$' indicates a representative point in the CE region
for which the correlation functions are presented here.}
\end{figure}
%%%%%%%%%%%%%%%%%%%%%%%%%%%%%%%%%%%%%%%%%%%%%%%%%%%%%%%%%%%%%%%%%%%%%%%%%%%%

%%%%%%%%%%%%%%%%%%%%%%%%%%%%%%%% FIG. 8 %%%%%%%%%%%%%%%%%%%%%%%%%%%%%%%%%%%%
\begin{figure}
\includegraphics[width=10cm]{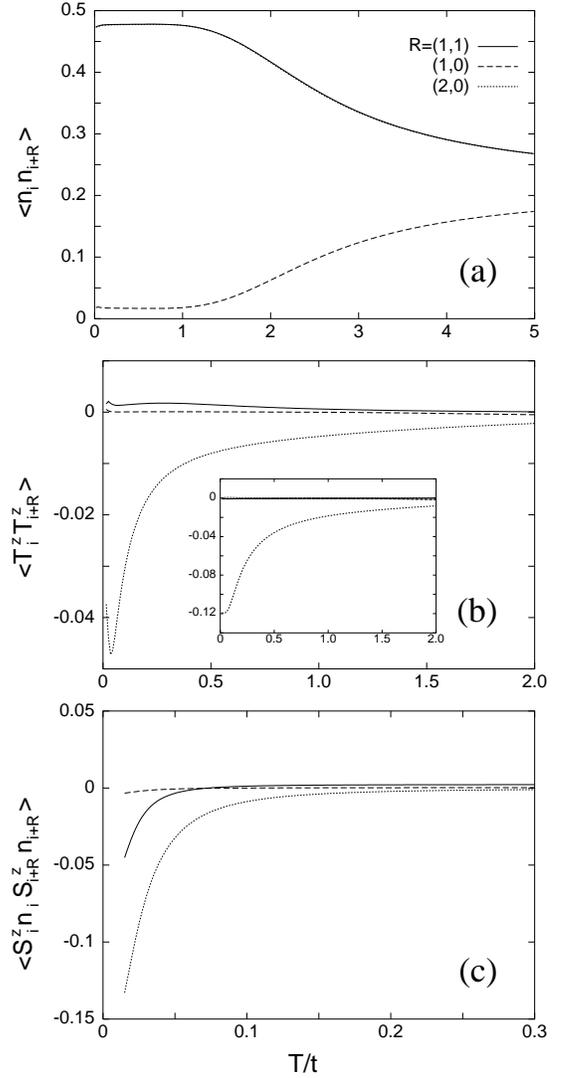}
\caption{\label{fig:corr_CE_el}
Temperature dependence of the two-site correlations as in 
Fig.~\protect{\ref{fig:corr_C_NN}} but for the phase with CE spin and orbital
correlations promoted by elastic interactions ($\kappa=0.2t$, 
$\kappa'=\kappa/8$). Other parameters: $V=3t$, $J_\textrm{AF}=0.04t$, 
$J_H=15t$.
The inset in (b) shows the orbital correlations in the rotated basis:
$\{(|x\rangle\pm|z\rangle)/\sqrt{2}\}$.}
\end{figure}
%%%%%%%%%%%%%%%%%%%%%%%%%%%%%%%%%%%%%%%%%%%%%%%%%%%%%%%%%%%%%%%%%%%%%%%%%%%%

The CE phase is favored as compared to the  C phase 
by the OO coupling between second neighbors ($\sim \kappa'$) 
along $a$/$b$ direction. However, this interaction is only active 
(not screened) when strong charge ordering takes place (large $V/t$ case) 
and here the CE phase can persist even for small values of 
$J_\textrm{AF}$ as long as the $e_g$ 
kinetic energy is sufficiently suppressed by a large $V$ 
(see Fig.~\ref{fig:elastic_phase}).
Thus, one has to assume large inter-site Coulomb repulsion (where almost 
perfect Mn$^{+3}$/Mn$^{+4}$ charge ordering\cite{Mur98} is realized) 
and small superexchange interaction [see an example for $V/t=3$ and 
$J_\textrm{AF}/t=0.04$ in Fig.~\ref{fig:corr_CE_el}].
In the $\{|x\rangle,|z\rangle\}$ basis, which we use in most of the numerical
evaluations, the spin and orbital correlation functions with 
${\bf R}=(2,0)$ become large
as $T\to 0$ while all the others remain very small in the whole 
range of temperatures. Such small orbital correlations imply either that the 
respective orbitals are uncorrelated or that they order in such a way, 
that the occupied orbitals are linear combinations with almost equal
amplitude expressed in the basis used for the measurement of correlation 
functions.\cite{Hor99} This difficulty is not present in SU(2) symmetric
models, and appears here as consequence of the cubic symmetry in the orbital 
sector.  
Therefore, the same functions were evaluated also 
in the $\{(|x\rangle\pm|z\rangle)/\sqrt{2}\}$ basis as presented in 
the inset of Fig.~\ref{fig:corr_CE_el} (b). In the new orbital basis the 
function $\langle T^z_{\bf i}T^z_{{\bf i}+(2,0)}\rangle$ is more pronounced 
and saturates for $T/t\simeq 0.1$ at 
$\langle T^z_{\bf i}T^z_{{\bf i}+(2,0)}\rangle\simeq -0.12$ close to its 
maximum value $-1/8$ indicating almost perfect CE orbital 
structure (see Fig.~\ref{fig:ce_orb}). Shorter range 
$\langle T^z_{\bf i}T^z_{{\bf i}+{\bf R}}\rangle$
correlations should become small when averaged over differently 
oriented clusters (rotated by angles $\pm\pi/2$) [see 
Fig.~\ref{fig:spins} (d)]. From
Fig.~\ref{fig:spins} (d) one can see that a similar pattern as found for 
the two-site orbital correlations should be found for the
$\langle S^z_{\bf i}n_{\bf i}S^z_{{\bf i}+{\bf R}}n_{{\bf i}+{\bf R}}\rangle$
functions. In our cluster in ideal CE state 
only electrons at distance $|{\bf R}|=2$ can contribute 
to the charge-spin correlations (being negative in this case) which
do not vanish when averaged over different cluster orientations. 
As shown in Fig.~\ref{fig:corr_CE_el} (c) the 
$\langle S^z_{\bf i}n_{\bf i}S^z_{{\bf i}+{\bf R}}n_{{\bf i}+{\bf R}}\rangle$
correlations obtained here are 
consistent with the above characterization of the CE pattern with a strong 
negative signal dominating at distance $|{\bf R}|=2$. The existence of
charge-ordered insulating state concomitant with orbital and spin order
of the CE-type\cite{Wol55,Goo55} was observed in different half-doped 
manganites with small one-electron bandwidth.\cite{Ima98}

\begin{figure}
\includegraphics[width=6cm]{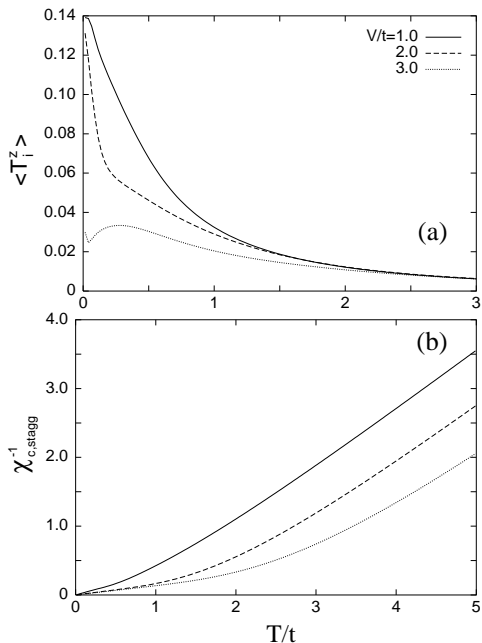}
\caption{\label{fig:Tz_susc_el}
Temperature dependence of the average value of the $T_i^z$
orbital operator (a) and the inverse of the staggered charge susceptibility 
$\chi^{-1}_{c,stagg}$ (b) obtained for different values of the inter-site 
repulsion $V/t$ in the presence of elastic interactions ($\kappa=0.2t$, 
$\kappa'=\kappa/8$). Other parameters: $J_\textrm{AF}=0.04t$, $J_H=15t$.}
\end{figure}

The average change in the orbital occupation can be studied calculating
$\langle T^z_{\bf i}\rangle$ which changes between $+1/4$ and $-1/4$ 
for the uniform $|x\rangle$ and $|z\rangle$ orbital states, respectively,
for the case with one $e_g$ electron per two Mn sites.
Furthermore, when the orbital state involves $|x\rangle$ and $|z\rangle$ 
orbitals in the same proportion one finds 
$\langle T^z_{\bf i}\rangle\simeq 0$. In Fig.~\ref{fig:Tz_susc_el} (a) we see
that increasing the Coulomb interaction from $V=t$ to $V=2t$ the orbital
ordering hardly changes for $T/t\to 0$ with $\langle T^z_{\bf i}\rangle$
only slightly decreasing to $\langle T^z_{\bf i}\rangle\simeq 0.13$ for
$V=2t$. This indicates that as long as the system is in the FM state 
(see Fig.~\ref{fig:elastic_phase}) the ground state is dominated by 
the $x^2-y^2$
orbital promoted by the kinetic energy term. Further increase in the Coulomb
inter-site repulsion to $V=3t$ leads to a drastic change in the orbital order
when system enters the CE phase region (see Fig.~\ref{fig:elastic_phase}). 
Here, 
$\langle T^z_{\bf i}\rangle\simeq 0.03$ for $T/t\to 0$ indicating dominating 
role played by the $(|x\rangle\pm|z\rangle)/\sqrt{2}$ orbitals. This fact
is consistent with our results for the two-site orbital correlation functions
[see inset in Fig.~\ref{fig:corr_CE_el}(b)].

%%%%%%%%%%%%%%%%%%%%%%%%%%%%%%%% FIG. 9 %%%%%%%%%%%%%%%%%%%%%%%%%%%%%%%%%%%%
%%%%%%%%%%%%%%%%%%%%%%%%%%%%%%%%%%%%%%%%%%%%%%%%%%%%%%%%%%%%%%%%%%%%%%%%%%%%

A change in the inter-site Coulomb interaction is directly connected with
the robustness of the charge ordering as seen in 
Figs. \ref{fig:corr_FM_NN} - \ref{fig:corr_C_NN} (a) and 
Fig.~\ref{fig:corr_CE_el} (a). 
Although the characteristic temperature of the CO ($T_\textrm{CO}$)
can be roughly estimated from the charge-charge correlation functions 
it can be more precisely extracted from the staggered charge 
susceptibility,\cite{Mack99} $\chi_{c,stagg}$:
\begin{equation}
\label{eq:susc}
\chi_{c,stagg} = \beta/N\sum_{\bf ij}
(-1)^{|{\bf i}-{\bf j}|}\langle\left(n_{\bf i} - \frac{1}{2}\right)
\left(n_{\bf j} - \frac{1}{2}\right)\rangle ,
\end{equation}
which at high temperatures follows a Curie-Weiss law 
$\chi_{c,stagg}^{-1}\propto (T - T_\textrm{CO})$. 
From Fig.~\ref{fig:Tz_susc_el} (b)
one can easily estimate $T_\textrm{CO}/t\approx 0.6$, $1.4$, and $2.4$ for 
$V/t=1.0$, $2.0$, and $3.0$, respectively, in agreement with the mean-field
solution\cite{Yu00} which for $V\gg t$ predicts, $T_\textrm{CO}\approx zV/4$.

%%%%%%%%%%%%%%%%%%%%%%%%%%%%%%%%%%%%%%%%%%%%%%%%%%%%%%%%%%%%%%%%%%%%%%%%%%%
\subsection{\label{sec:nnnjt} Microscopic model for Mn$^{3+}$--Mn$^{3+}$ 
JT coupling}
%%%%%%%%%%%%%%%%%%%%%%%%%%%%%%%%%%%%%%%%%%%%%%%%%%%%%%%%%%%%%%%%%%%%%%%%%%%%

Although the model considered in the previous section with longer-range OO 
elastic coupling leads to the CE-type
spin and orbital correlations, this phase is found only in a very extreme 
case of almost perfect Mn$^{4+}$/Mn$^{3+}$ CO (that is for strong Coulomb 
repulsion) which may not be very realistic in 
the light of some experimental facts indicating the CDW character of charge 
redistribution between different manganese ions.\cite{Lar01} 
Therefore in this section we derive the further neighbor OO interactions 
from a microscopic model including apart from the JT-distortion of the 
Mn$^{3+}$ octahedra the breathing distortion around the Mn$^{4+}$ ions
as well.
We then show that this interaction is capable to produce
the zigzag (spin and orbital) phase in a region of small Coulomb repulsion 
changing the previous phase diagrams in a drastic way.  

\begin{figure}
\includegraphics[width=7.5cm]{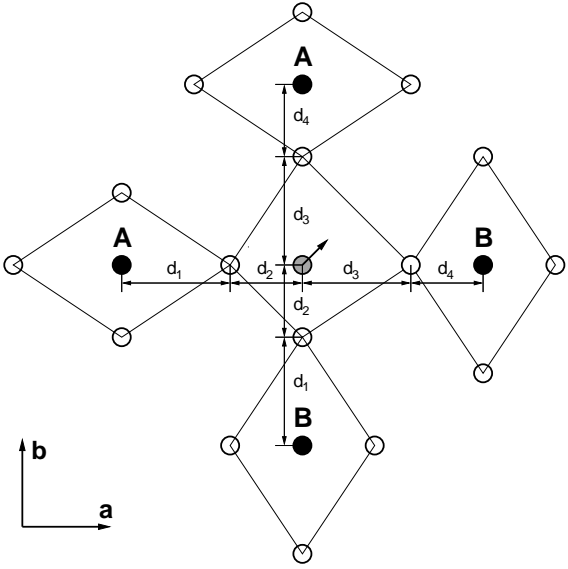}
\caption{\label{fig:distortions}
Schematic representation of the lattice distortions contributing to the
second neighbor JT-interaction. The corresponding CE orbital order for  the
5 atom structure above is indicated 
in Fig.~\protect{\ref{fig:ce_orb}} by dashed lines. 
Different Mn$^{3+}$, Mn$^{4+}$, and O ions are presented by full, shaded 
and empty circles, respectively. The distances between the Mn$^{4+}$ ion
and the Mn$^{3+}$ ions in $(10)$ [$(\bar{1}0)$] and $(01)$ [$(0\bar{1})$]
directions, respectively, are identical. The arrow indicates the displacement 
of the central Mn$^{4+}$ ion when $d_1+d_2\neq d_3+d_4$.}
\end{figure}

To assess the gain in lattice energy connected with the formation 
of the CE long-range orbital order we follow the work of Millis for LaMnO$_3$
by treating the lattice distortions classically.\cite{Mil96,Ahn98}
The effect of lattice displacements is described by assuming arbitrary 
values of four different lattice distances $d_i$ ($i=1,...,4$) 
(see Fig.~\ref{fig:distortions}) which are expressed by 
$\delta_x$, $\delta'_x$ -- uniform 
deformation along the $a$ and $b$ direction of the Mn lattice, 
and $u_i$ -- displacements of O ions along the Mn--O--Mn 
bonds within the $(a,b)$ planes:
\begin{subequations}
\begin{eqnarray}
d_1 &=& \frac{1}{2}b\left(1 + \delta_x - 2u_1\right) ,\\
d_2 &=& \frac{1}{2}b\left(1 + \delta_x + 2u_1\right) ,\\
d_3 &=& \frac{1}{2}b\left(1 + \delta'_x + 2u_2\right) ,\\
d_4 &=& \frac{1}{2}b\left(1 + \delta'_x - 2u_2\right) ,
\end{eqnarray}
\end{subequations}
where $b$ is the lattice constant of the ideal perovskite.\cite{noncentral} 
Moreover, we assume $e_g$ charge stacking in the $c$ direction with 
the respective Mn--Mn distance being $b(1+\delta_z)$.

%%%%%%%%%%%%%%%%%%%%%%%%%%%%%%%% FIG. 10 %%%%%%%%%%%%%%%%%%%%%%%%%%%%%%%%%%%%
%%%%%%%%%%%%%%%%%%%%%%%%%%%%%%%%%%%%%%%%%%%%%%%%%%%%%%%%%%%%%%%%%%%%%%%%%%%%

%%%%%%%%%%%%%%%%%%%%%%%%%%%%%%%% FIG. 11 %%%%%%%%%%%%%%%%%%%%%%%%%%%%%%%%%%%
\begin{figure}
\includegraphics[width=3.5cm]{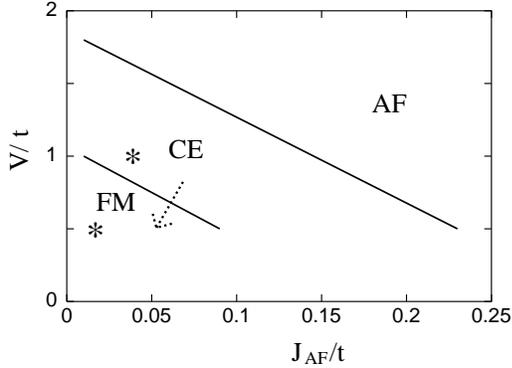}
\caption{\label{fig:OONNN_phase}
Phase diagram as in Fig.~\protect{\ref{fig:OONN_phase}} including
the local nearest- and next-neighbor OO interactions with $\kappa=0.2t$ and
$\kappa'=0.1t$ (spring constants $K_2/K_1=0.25$). 
'$*$' indicate representative points in 
the CE region for which the correlation functions are presented here.
The dashed arrow presents the direction of the transition from the localized
CE to the itinerant electron A-phase at low temperatures.}
\end{figure}
%%%%%%%%%%%%%%%%%%%%%%%%%%%%%%%%%%%%%%%%%%%%%%%%%%%%%%%%%%%%%%%%%%%%%%%%%%%%

In order to define
orbital sublattices (A and B)  we introduce a transformation which describes 
the tilting of orbitals by making two different transformations at both 
sublattices,\cite{Bri99}
\begin{equation}
\label{eq:rotation}
\left[
\begin{array}{c}
|i\mu\rangle\\
|i\nu\rangle
\end{array}\right] =
\left[
\begin{array}{cc}
\cos\left(\frac{\pi}{4}\pm\phi\right) 
& \sin\left(\frac{\pi}{4}\pm\phi\right) \\
-\sin\left(\frac{\pi}{4}\pm\phi\right)
& \cos\left(\frac{\pi}{4}\pm\phi\right)
\end{array}\right] 
\left[
\begin{array}{c}
|i z\rangle\\
|i x\rangle
\end{array}\right] ,
\end{equation}
where $-$ ($+$) refers to $i\in A$ ($i\in B$) sublattice, respectively,
and $|\phi|\leq\frac{\pi}{4}$.
Here, $|ix\rangle$ and $|iz\rangle$ stands for local basis orbitals:
$|x\rangle$ and $|z\rangle$ at site $i$, respectively.
In the rotated basis $\{|i\mu\rangle,|i\nu\rangle\}$ we assume for 
the transformed operators: $\langle T^z_i\rangle = -1/2$ ($1/2$) for 
$i\in A$ ($B$) sublattice, respectively. At intermediate values of $\phi$ 
one can reach, e.g., $|3x^2-r^2\rangle$/$|3y^2-r^2\rangle$
($|z^2-x^2\rangle$/$|z^2-y^2\rangle$)-type orbital ordering at 
$\phi=\pi/12$ ($\phi=-\pi/12$), respectively. The rotation (\ref{eq:rotation})
does not include complex orbital order which was proposed as a possible
orbital order in manganites at smaller doping.\cite{complex}
The classical energy of the distorted lattice
$E_l$ (per Mn site) is given by,\cite{Ahn98}
\begin{equation}
\label{eq:E_l}
E_l= 
\frac{1}{4}\left(K_1 + 2K_2\right)
\left[\delta_x^2 + (\delta'_x)^2 + \delta_z^2\right]
+ K_1\left(u_1^2 + u_2^2\right),
\end{equation}
normalized per Mn ion, where $K_1$ ($K_2$) are the nearest-neighbor Mn--O 
(Mn--Mn) spring constant, respectively. Here the force constants include
the factor $b^2$, and hence have the dimension of an energy.

Next, we have to add the energy of the breathing mode (BM) describing 
the attraction between oxygen ions and the unoccupied Mn$^{+4}$ site:
\begin{equation}
E_{br} = \frac{1}{2}\beta\lambda\left(\delta_x + \delta'_x + \delta_z +
u_1 + u_2\right) .
\label{eq:br}
\end{equation}
Here, all oxygens are equally attracted towards the central manganese ion.
The BM induces charge alternation on neighboring Mn sites and contributes to 
the effective inter-site charge-charge repulsion ($V$). The coupling constant
to the breathing distortion is written as $\beta \lambda$, where $\lambda$
denotes the coupling to the JT distortion. 
The term describing the JT coupling depends on the character
of the occupied orbital at a Mn$^{+3}$ site.
The JT interaction describes the coupling between the occupied 
$e_g$ orbital at 
a particular Mn$^{3+}$ ion and the distortions of the surrounding oxygen 
ions, as introduced by Millis.\cite{Mil96} In the orbitally rotated state 
(\ref{eq:rotation}) this energy has the following form,
\begin{eqnarray}
\label{eq:JT}
E_\textrm{JT}(\phi) &=& \frac{1}{2}\lambda\left\{\left[\delta_z + u_1 + u_2 - 
\frac{1}{2}\left(\delta_x + \delta'_x\right)\right]\sin(2\phi)\right.
\nonumber\\
&-& \left.\sqrt{3}\cos(2\phi)\left[\frac{1}{2}\left(\delta_x 
- \delta'_x\right) + u_2 - u_1\right]\right\},
\end{eqnarray}
and depends on the rotation angle $\phi$ and on the deformation 
of the lattice.

The optimal values of the ionic displacements for a given angle 
$\phi$ are found 
by minimization of the total energy, 
$E_{tot}(\phi)=E_l+E_{br}+E_\textrm{JT}(\phi)$, 
which for (\ref{eq:E_l})-(\ref{eq:JT}) and neglecting higher-order nonlinear 
terms gives,
\begin{subequations}
\label{eq:dis}
\begin{eqnarray}
\delta_x &=& \frac{-\lambda\left[2\beta-\sin(2\phi)-\sqrt{3}\cos(2\phi)\right]}
{2\left(K_1+2K_2\right)} ,
\\
\delta'_x &=& \frac{-\lambda\left[2\beta
-\sin(2\phi)+\sqrt{3}\cos(2\phi)\right]}
{2\left(K_1+2K_2\right)} ,
\\
\delta_z &=& \frac{-\lambda\left[\beta+\sin(2\phi)\right]}{K_1+2K_2} ,
\\
u_1 &=& \frac{-\lambda\left[2\beta+\sin(2\phi)+\sqrt{3}\cos(2\phi)\right]}
{4K_1} ,
\\
u_2 &=& \frac{-\lambda\left[2\beta+\sin(2\phi)-\sqrt{3}\cos(2\phi)\right]}
{4K_1} .
\end{eqnarray}
\end{subequations}
This yields for the total energy of the distorted lattice,
\begin{eqnarray}
\label{eq:E_tot}
&&E_{tot}(\phi) = -\frac{\lambda^2}{4K_1(K_1+2K_2)}\left\{
\beta^2(5K_1+4K_2)\right.\nonumber \\
&+& \left.(K_1+2K_2)\left[2\beta\sin(2\phi)
+\cos^2(2\phi)\right]+2K_1+K_2\right\} .\nonumber \\
\end{eqnarray}
This implies a change of the volume of the lattice with respect to 
the ideal perovskite ($V_0=b^3$)  $\delta V/V_0\simeq 
\delta_x+\delta'_x+\delta_z = -3\lambda\beta/(K_1+2K_2)$. Hence
the volume decreases with increasing BM coupling $\beta$, 
but does not depend on the character of the orbital order.

The relative strength of the effective next- and nearest-neighbor JT 
interactions can be obtained by comparing the lattice stiffnesses of the 
orbital modulations in the undoped ($\sim\kappa$) and the
quarter-filled ($\sim\kappa'$) case, respectively. 
For \textit{undoped} LaMnO$_3$ the lattice 
energy was found to be [see Eq.~(2.14) in Ref.~\onlinecite{Bal02}],
\begin{equation}
E_{tot}^0(\phi)=-\frac{3\lambda^2}{K_1(K_1+2K_2)}\left\{K_1+K_2
\left[1 + \cos(4\phi)\right]\right\},
\label{eq:E^0}
\end{equation}
In the half-doped case the orbital alternation at Mn$^{3+}$ neighbors
is driven by the Mn$^{4+}$ ion shift.
The energy (21) can be decomposed $E_{tot}(\phi)=E_0(\phi)+\delta E(\phi)$,
where $\delta E(\phi)$ is the gain of energy due to the Mn$^{4+}$
displacement:
\begin{equation}
\delta E(\phi)=-\frac{3\lambda^2}{16(K_1+2K_2)}\left[1 + \cos(4\phi)
\right].
\label{eq:dE}
\end{equation}
This cooperative energy does not depend on the breathing mode coupling
$\beta$ directly, though there is an indirect dependence via $E_0(\phi)$ 
influencing the optimal value for $\phi$.
Contrary to the undoped case (\ref{eq:E^0}), where the Mn-Mn potential 
($\sim K_2$) is necessary to obtain finite orbital stiffness, 
in the half-doped case (\ref{eq:dE}) the Mn-O interaction alone can lead to
cooperative JT effect stabilizing the alternating orbital order. Assuming 
$\lambda=6$eV, $K_1=200$ eV, 
and $0<K_2/K_1<1$ (see Ref.~\onlinecite{Mil96,dimension})
one finds for the $|x\rangle\pm|z\rangle$ ($\phi=0$) orbital order
the CE energy contribution per Mn ion $\delta E\simeq 20 - 70$ meV.

In Table \ref{tab:ene} we present the ratio $\delta E(\phi)/E_{tot}(\phi)$,
characterizing the stability of the CE orbital order, as function of the BM 
coupling strength $\beta$
and the ratio of spring constants $K_2/K_1$. 
The largest relative gain in energy due to 
the Mn$^{4+}$ ions shift ($\sim |\delta_x-\delta'_x|b/2$) is obtained for 
the $(|x\rangle+|z\rangle)/(|x\rangle-|z\rangle)$ orbital order ($\phi=0$) 
which for small BM coupling ($\sim\beta$) can reach up to $50\%$ of the total 
lattice energy, whereas for larger values of $\beta$, although 
$\delta E(\phi)$ is not changed, the total lattice energy 
increases rapidly as $\sim \beta^2$ being dominated by 
the distortions of the MnO$_6$ octahedra rather then by their displacements. 
Furthermore, with increasing Mn lattice stiffness ($\sim K_2/K_1$) 
both the $\delta E(\phi)/E_{tot}(\phi)$ ratio and the shift of unoccupied 
Mn$^{4+}$ ions [$\sim \lambda/(K_1+2K_2)$] decrease.

%%%%%%%%%%%%%%%%%%%%%%%%%%%%%%%% TAB. 1 %%%%%%%%%%%%%%%%%%%%%%%%%%%%%%%%%%%%
\begin{table}
\caption{\label{tab:ene}
The relative energy gain, $\delta E(\phi)/E_{tot}(\phi)$, 
associated with the Mn$^{4+}$ ion displacement (see Fig.
\protect{\ref{fig:distortions}}) for the orbital orderings given by 
$\phi=0$ and $\pm\pi/12$ presented as function of $\beta$ and $K_2/K_1$.}
\begin{ruledtabular}
\begin{tabular}{lcccc}
& & $\phi=\pi/12$ & $\phi=0$ & $\phi=-\pi/12$ \\
\hline
$\beta$ & $K_2/K_1$ & $\delta E(\phi)/E_{tot}(\phi)$ 
& $\delta E(\phi)/E_{tot}(\phi)$ 
& $\delta E(\phi)/E_{tot}(\phi)$ \\
\hline
0.0 & 0.0 & 0.409 & 0.500 & 0.409  \\ 
0.5 & 0.0 & 0.321 & 0.353 & 0.250  \\ 
1.0 & 0.0 & 0.167 & 0.187 & 0.129  \\ 
\hline
0.0 & 0.5 & 0.281 & 0.333 & 0.281  \\ 
0.5 & 0.5 & 0.237 & 0.240 & 0.167  \\ 
1.0 & 0.5 & 0.125 & 0.130 & 0.087  \\ 
\hline 
0.0 & 1.0 & 0.214 & 0.250 & 0.214  \\ 
0.5 & 1.0 & 0.187 & 0.182 & 0.125  \\ 
1.0 & 1.0 & 0.100 & 0.100 & 0.065  \\ 
\end{tabular}
\end{ruledtabular}
\end{table}
%%%%%%%%%%%%%%%%%%%%%%%%%%%%%%%%%%%%%%%%%%%%%%%%%%%%%%%%%%%%%%%%%%%%%%%%%%%%

%%%%%%%%%%%%%%%%%%%%%%%%%%%%%%%% FIG. 12 %%%%%%%%%%%%%%%%%%%%%%%%%%%%%%%%%%%
\begin{figure}
\includegraphics[width=10cm]{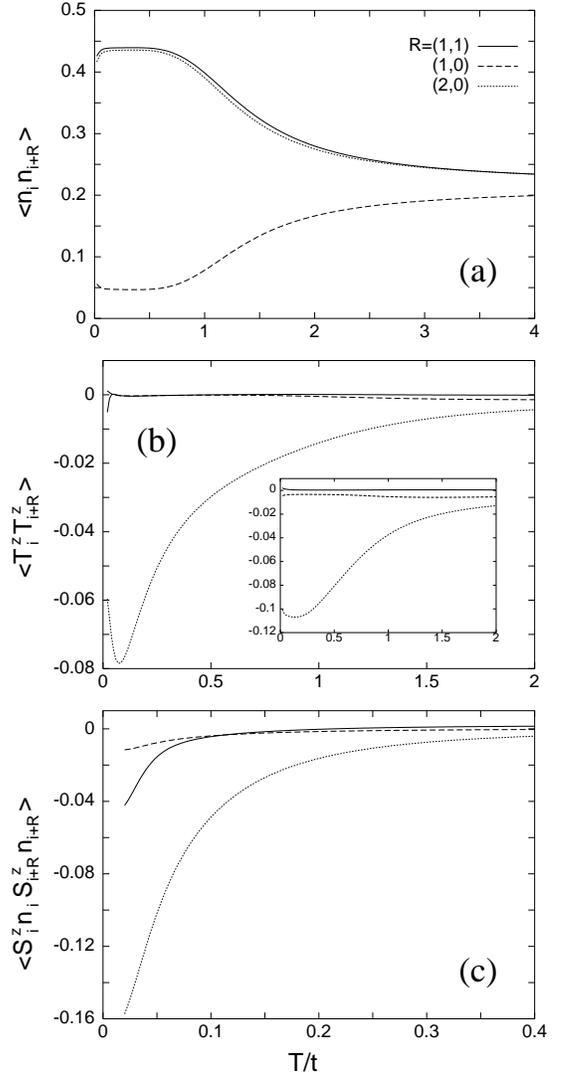}
\caption{\label{fig:corr_CE_NNN}
Correlation functions in the case of a CE ground state (as in 
Fig.~\protect{\ref{fig:corr_CE_el}}) but obtained for the local OO interaction 
(26) with spring constants $K_2/K_1=0.25$ ($\kappa'=\kappa/2$). 
Other parameters: $V=t$, $\kappa=0.2t$, $J_\textrm{AF}=0.04t$, $J_H=15t$.}
\end{figure}
%%%%%%%%%%%%%%%%%%%%%%%%%%%%%%%%%%%%%%%%%%%%%%%%%%%%%%%%%%%%%%%%%%%%%%%%%%%%

%%%%%%%%%%%%%%%%%%%%%%%%%%%%%%%% FIG. 13 %%%%%%%%%%%%%%%%%%%%%%%%%%%%%%%%%%%
\begin{figure}
\includegraphics[width=3.5cm]{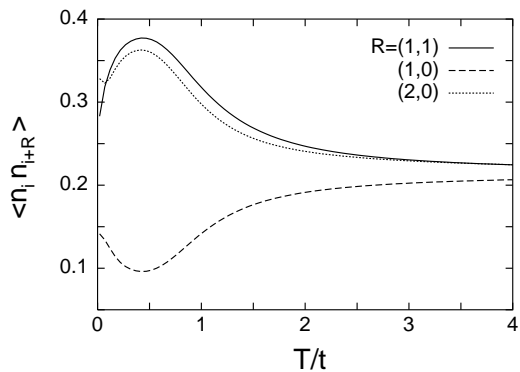}
\caption{\label{fig:<nn>ce}
Charge correlations in the case of a FM ground state as 
obtained for the local OO interaction 
with spring constants $K_2/K_1=0.25$ ($\kappa'=\kappa/2$). 
Other parameters: $V=0.5t$, $\kappa=0.2t$, $J_\textrm{AF}=0.02t$, $J_H=15t$.}
\end{figure}
%%%%%%%%%%%%%%%%%%%%%%%%%%%%%%%%%%%%%%%%%%%%%%%%%%%%%%%%%%%%%%%%%%%%%%%%%%%%

Comparing the cooperative part [$\sim\cos(4\phi)$] of both JT energy 
contributions [(\ref{eq:E^0}) and (\ref{eq:dE})] we end up with a simple 
relation between different OO interaction strengths,
\begin{equation}
{\frac{\kappa'}{\kappa}}={\frac{K_1}{8K_2}} .
\label{eq:kappa_ratio}
\end{equation}
For $K_2=K_1$ the $\sim R^{-3}$ scaling of  elastic interactions\cite{Kho01} 
is recovered. In the above estimate we have neglected 
the superexchange interaction which can also contribute to the effective 
$\kappa$ [see \textit{e.g.} Eq.~(2.7) in Ref.~\onlinecite{Bal02}].

The relative strength of both spring constants in the CO case can be assessed
from the relative length of different bonds in the half-doped MnO$_2$ plane.
As measured for low-temperature La$_{0.5}$Ca$_{0.5}$MnO$_3$ 
superstructure\cite{Rad97} the length of all four in-plane Mn$^{4+}$-O bonds
is very similar ($d_2=d_3\simeq 1.915\mathring{A}$). Neglecting the effect of 
Mn-O-Mn bond bending and assuming $d_2\simeq d_3$ (see 
Fig.~\ref{fig:distortions}) one obtains the  relation,
\begin{equation}
K_2\ll K_1 ,
\label{eq:K_ratio}
\end{equation}
by using Eqs.~(\ref{eq:dis}). The above assessment together with 
(\ref{eq:kappa_ratio}) underlines the importance of Mn$^{3+}$--Mn$^{3+}$ 
orbital coupling ($\sim\kappa'$) in CO manganites.

The structural data of Radaelli {\it et al.}\cite{Rad97} may be 
used in combination
with Eqs. (\ref{eq:dis} a-e) to infer the values for $\lambda$ and $\beta$.
We have not attempted to determine the force constants as well, 
but assume \cite{Mil96}
$K_1=200$ eV . Using basically data for the JT-distortion 
of Mn$^{3+}$--O in-plane bonds 
($d_1\simeq 2.07\mathring{A}$ and $d_4\simeq 1.92\mathring{A}$) and the
Mn$^{4+}$ displacement, 
we obtain for $K_2=0.25 (0.0) K_1$ $\beta\simeq 0.27 (0.35)$
and $\lambda\simeq 7.4 (5.7)$ eV, respectively. We note that the ratio
$\lambda/K_1\simeq 0.037 (0.029)$ is rather close to the values determined by 
Millis\cite{Mil96} for LaMnO$_3$.
This also implies that the energy gain due to the breathing distortion is
substantial [see Eq.~(\ref{eq:E_tot})].

%%%%%%%%%%%%%%%%%%%%%%%%%%%%%%%%%%%%%%%%%%%%%%%%%%%%%%%%%%%%%%%%%%%%%%%%%%%
\subsection{\label{sec:nnnjt0}Effect of cooperative Mn$^{3+}$--Mn$^{3+}$ 
JT interactions}
%%%%%%%%%%%%%%%%%%%%%%%%%%%%%%%%%%%%%%%%%%%%%%%%%%%%%%%%%%%%%%%%%%%%%%%%%%%%

We return to the many-body theory for a translation invariant system.
As the OO interaction between further Mn$^{3+}$ neighbors is mediated by 
an Mn$^{4+}$ ion 
we introduce the following effective three-site interaction,
\begin{equation}
H'_\textrm{OO}=2\kappa'\sum_{\langle{\bf ijj'}
\rangle}(1 - n_{\bf j})T_{\bf ij'} ,
\label{eq:3site}
\end{equation}
where the operator $T_{\bf ij'}$ is defined as in Eq.~(\ref{eq:Tij}), and
$\langle{\bf ijj'}\rangle$ denotes three neighboring sites along
$a$ or $b$ direction. The reasoning here is similar as for 
the nearest-neighbor JT
interaction in Eq.~(\ref{eq:H_OO}), i.e., the expectation value of the operator
$T_{\bf ij'}$
 in an orbital ordered state depends on the orbital angle 
precisely as in Eq.~(\ref{eq:dE}), namely
$\sim [1 + \cos(4\phi)]$. The $(1 - n_{\bf j})$ 
factor in Eq.~(\ref{eq:3site}) 
reflects 
the charge state of Mn$^{4+}$ and favors CO in the quarter-filled system, 
while the contribution of the interaction (\ref{eq:3site}) would vanish in 
the undoped compound.

Investigating the model (\ref{eq:H_total}) with both nearest-neighbor 
(\ref{eq:H_OO}) and three-site (\ref{eq:3site}) OO couplings we find a strong
change in the low-temperature phase diagram (see Fig.~\ref{fig:OONNN_phase})
as compared with the previous ones (Figs. \ref{fig:OONN_phase} and 
\ref{fig:elastic_phase}). Assuming $\kappa'=\kappa/2$ 
(which corresponds to $K_2/K_1=1/4$), the CE-like correlations set in 
already at rather small inter-site repulsion ($V\simeq 0.5t$) while for 
stronger Coulomb interactions ($V\agt 1.5t$) the spin correlations 
have predominantly AF character. 
Furthermore, the FM region is quite small and can be found 
only for $V\alt t$ and 
$J_\textrm{AF}\alt 0.1t$. When decreasing the ratio $K_2/K_1$ 
(and keeping the value of 
$\kappa'$ constant) the FM region shrinks at the expense of the CE region. 

Next, we present the two-site correlations for one representative 
($V=t$, $J_\textrm{AF}=0.04t$) point in the CE region in 
Fig.~\ref{fig:OONNN_phase}. From Fig.~\ref{fig:corr_CE_NNN} (a) one sees that 
the CE state can exist for less pronounced CO than in the elastic
strain model  considered in Sec.~\ref{sec:elastic}, where almost perfect 
Mn$^{3+}$/Mn$^{4+}$ CO was necessary to stabilize the zigzag state 
(see Fig.~\ref{fig:corr_CE_el}). 
Moreover, the evolution of CE-type orbital order with decreasing temperature
is here correlated with the onset of CO [compare Fig.~\ref{fig:corr_CE_NNN} 
(a)-(b) with \ref{fig:corr_CE_el} (a)-(b)]. Regarding 
$\langle S^z_{\bf i}n_{\bf i}S^z_{{\bf i}+{\bf R}}n_{{\bf i}+{\bf R}}\rangle$
correlations, the CE-type spin order develops at lower temperatures than the
charge and orbital order but at higher temperature
than previously found for the strain 
induced interactions. 
The three-site term (\ref{eq:3site}) which is active for a 
Mn$^{3+}$--Mn$^{4+}$--Mn$^{3+}$  sequence along the $a$ and $b$ direction,
respectively,
promotes not only orbital alternation on neighboring Mn$^{3+}$ sites but also
strengthens CO. This is clearly seen in the charge correlations which are now
stronger than those \textit{e.g.} in Fig.~\ref{fig:corr_FM_NN}, 
although the same Coulomb repulsion  ($V=t$) was assumed in both cases.

Going from the CE to the FM phase (as indicated by an arrow in 
Fig.~\ref{fig:OONNN_phase}) one finds that  the CDW 
is now reduced at low temperatures
in the FM state (Fig.~\ref{fig:<nn>ce}), while the charge modulation
is still  rather strong at 
higher temperatures ($T/t\approx 0.5$) above the N\'eel temperature where
the electron kinetic energy is quenched due to the random 
$t_{2g}$ spin orientations. 
This behavior is a clear manifestation of the
DE mechanism. 
We note that
the CE-type orbital correlations are still
present in the FM phase in Fig.~\ref{fig:OONNN_phase} for $V\agt 0.5t$,
whereas for smaller values for $V$ $x^2$-$y^2$ orbital correlations
dominate, which are characteristic for the metallic A-phase.

%%%%%%%%%%%%%%%%%%%%%%%%%%%%%%%%%%%%%%%%%%%%%%%%%%%%%%%%%%%%%%%%%%%%%%%%%
\subsection{\label{sec:stacking}Charge modulation in the $c$ direction}
%%%%%%%%%%%%%%%%%%%%%%%%%%%%%%%%%%%%%%%%%%%%%%%%%%%%%%%%%%%%%%%%%%%%%%%%%

Finally, we consider possible mechanisms stabilizing the charge 
stacking of CE-ordered planes \cite{Ari02} 
as observed in most charge-ordered manganites at doping $x=1/2$. 
Any interaction stabilizing such a stacking must overcome 
the inter-site charge-charge repulsion. There are two interactions which 
contribute to such a repulsion: (i) the electron-electron Coulomb force $V$; 
and (ii) the BM coupling. Both interactions will effectively tend to 
maximize the number of Mn$^{3+}$--Mn$^{4+}$ bonds. 
The BM coupling contribution to the inter-site charge-charge repulsion 
($V_\textrm{BM}$) can be estimated comparing 
the lattice energies with charge alternating and stacked along 
the $c$ direction 
which leads to the estimate $V_\textrm{BM}=\lambda^2[\beta-sin(2\phi)]^2
/(2K_1)$. Assuming typical values for
$\lambda=10t$ and $K_1=500t$ (see Ref.~\onlinecite{Mil96}) together 
with our previous estimate $\beta\simeq 0.35$ one can obtain \textit{e.g.}
$V_\textrm{BM}\simeq 0.012t$ for the $\{(|x\rangle\pm|z\rangle)/\sqrt{2}\}$ 
CE orbital order. 

In a multi-layer system the OO interaction along the $c$ direction is given
by the nearest-neighbor coupling $\sim\kappa$, which in the 
$\{|x\rangle,|z\rangle\}$ basis has the form,\cite{Bri99}
\begin{equation}
H_\textrm{OO}^c=8\kappa\sum_{\langle{\bf ij}\rangle c}T^z_{\bf i}T^z_{\bf j} .
\end{equation}
%

%%%%%%%%%%%%%%%%%%%%%%%%%%%%%%%% FIG. 14 %%%%%%%%%%%%%%%%%%%%%%%%%%%%%%%%%%%
\begin{figure}
\includegraphics[width=7cm]{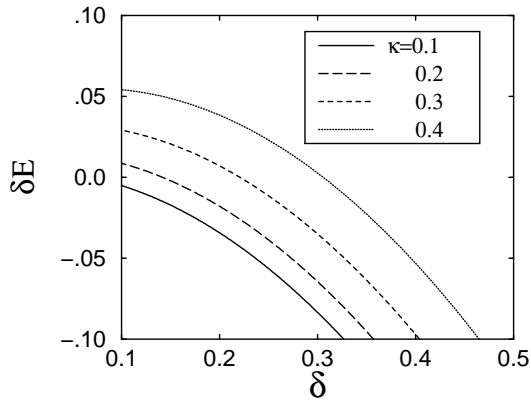}
\caption{\label{fig:stacking} 
Energy gained by $e_g$ electron stacked in the $c$ direction 
as function of charge modulation $\delta$ obtained
for $V=0.5t$, $\kappa'=0.1t$ and different values of $\kappa$.}
\end{figure}
%%%%%%%%%%%%%%%%%%%%%%%%%%%%%%%%%%%%%%%%%%%%%%%%%%%%%%%%%%%%%%%%%%%%%%%%%%%%

This interaction together with the $\sim\kappa'$ coupling, which controls 
the CE-orbital order in the $(a,b)$ plane, in a charge-ordered case yields 
an approximate orbital model of the form,
\begin{equation}
\label{eq:stacking}
H_\textrm{OO}\simeq 6\kappa'\sum_{{\langle\langle\bf ij\rangle\rangle}ab}
T^z_{\bf i}T^z_{\bf j}
+ 8\kappa\sum_{\langle{\bf ij}\rangle c}T^x_{\bf i}T^x_{\bf j} .
\end{equation}
where $T^z_{\bf i}$ ($T^x_{\bf i}$) are defined with respect to
the $\{(|x\rangle\pm|z\rangle)/\sqrt{2}\}$ orbital basis. 
Defining $|\pm\rangle\equiv(|x\rangle\pm|z\rangle)/\sqrt{2}$ and
assuming that in the $(a,b)$ planes of a bilayer system 
$|+\rangle_{\bf i}$ and $|+\rangle_{\bf j}$ orbitals are occupied on two 
neighboring sites in the $c$ direction, the $\sim\kappa$ coupling 
[see Eq.~(\ref{eq:stacking})] rotates orbitals towards the 
$|-\rangle_{\bf i}\otimes|-\rangle_{\bf j}$ configuration and leads to 
binding of two electrons with energy
$E_b=\sqrt{(12\kappa')^2+4\kappa^2}-12\kappa'$.
For charge disproportion with Mn$^{3.5\pm\delta}$ ions alternating in 
the $(a,b)$ planes and assuming that the $\kappa$ ($\kappa'$) 
two (three)-site interaction scales as $\sim(1/2+\delta)^\alpha$ 
with $\alpha=2$ ($3$), respectively, one finds for the energy gain (per 
Mn ion) due to charge stacking along $c$ direction for a 3D lattice:
\begin{equation}
\label{eq:deltaE}
\delta E=6\kappa'(\frac{1}{2}+\delta)^3
\left[\sqrt{1+\frac{\kappa^2}{36\kappa'^2(\frac{1}{2}+\delta)^2}}-1\right]
-2\delta^2V .
\end{equation}
The last term stands for the effective Coulomb repulsion at given
charge modulation. With charge modulation sufficiently reduced 
$\delta\sim 0.2-0.3$ (see Fig.~\ref{fig:stacking}) the OO inter-plane 
coupling can stabilize charge-stacked phase ($\delta E>0$). Such 
stabilization is strengthened (weakened) by increasing $\kappa$ ($\kappa'$)
OO interaction, respectively. For smaller charge disproportion ($\delta\alt
0.1$) and considerable charge on Mn$^{3.5-\delta}$ sites the $\sim\kappa$ 
in-plane coupling [neglected in Eq.~(\ref{eq:stacking})] 
can also play an important role in the inter-plane charge modulation.
However, when the CO is strongly reduced the orbital ground state is 
too complex to be analyzed analytically.

A further mechanism stabilizing the observed $e_g$ charge stacking along 
the $c$ direction results from the AF inter-plane 
superexchange interaction.\cite{Yun00} 
%which (in order to make all the bonds in the $c$ direction AF) 
%tends to stack the electrons on bridge sites of different zigzags.
Yet, the magnetic interaction ($J_{\rm AF}$) results mainly from 
the $t_{2g}-t_{2g}$ superexchange  which is of the order of only 
a few meV.\cite{Fei99} Thus, the antiferromagnetism in the $c$ direction
observed in most of the CE structures appears to be a \textit{result}
rather than the \textit{origin} of the stacking pattern of the charge/orbital
order.

%%%%%%%%%%%%%%%%%%%%%%%%%%%%%%%%%%%%%%%%%%%%%%%%%%%%%%%%%%%%%%%%%%%%%%%%%%%%
\section{\label{sec:summary}Summary}
%%%%%%%%%%%%%%%%%%%%%%%%%%%%%%%%%%%%%%%%%%%%%%%%%%%%%%%%%%%%%%%%%%%%%%%%%%%%

We have investigated the potential of different electron-lattice mechanisms  
to stabilize the CE spin-orbital ordered phase in charge ordered half-doped
manganites (filled with one $e_g$ electron per two Mn ions). Our study
is based on
a generalized DE model containing the $e_g$ orbital degrees of 
freedom, and also including inter-site Coulomb repulsion ($V$) and
AF superexchange ($J_\textrm{AF}$) between core spins. 
The effect of
three types of orbital interactions derived from the cooperative JT
effect have been studied:
(i) nearest neighbor JT-coupling, which is responsible for the alternating
orbital order in undoped LaMnO$_3$,
(ii) in addition second neighbor JT-interaction with a strength as expected
for elastic strain, and
(iii) a 2nd neighbor JT interaction which accounts for the breathing
distortion of the Mn$^{4+}$ octahedra and is in general
stronger than (ii). 
A particular feature of this interaction, 
which has been proposed based on the classical model considerations
in Sec.~\ref{sec:nnnjt}, are the displacements of the  Mn$^{4+}$ ions 
in the CE-phase. 

We have studied these models employing finite temperature diagonalization
on a $\sqrt{8}\times\sqrt{8}$ cluster with periodic boundary conditions, 
thereby
keeping the translational invariance. The evolution of the different phases 
with temperature is monitored by the calculation of correlation functions for 
charge, orbital and spin, respectively.
Starting out from the generalized FM Kondo lattice
model (\ref{eq:H_total}) without any orbital interaction
($H_\textrm{OO}=0$) we find the orbital correlations dominated by 
the $x^2-y^2$ 
orbitals which are favored by the kinetic energy.\cite{Hor99,Mac99}
This orbital state is characteristic for the metallic A-phase
in wide-band manganites at half doping.  
For small AF superexchange interactions ($J_\textrm{AF}\ll t$) between 
$t_{2g}$ core spins the FM state is found in the spin sector, while
for larger values of $J_\textrm{AF}$ the ground state is that of a regular 
2 sublattice antiferromagnet. Even for large Coulomb repulsion $V$ we
found no evidence for C or CE structures for some intermediate values
of the parameter  $J_\textrm{AF}$. 
The inclusion of the nearest-neighbor JT-interaction
(i), however, was found to stabilize the AF-C phase in a wide parameter
range in the $V-J_\textrm{AF}$ phase diagram.
 
To obtain the more complex CE orbital and spin state 
not only the kinetic energy must be reduced
by increasing the Coulomb repulsion $V$ but also a further neighbor 
\textit{antiferro-orbital} interaction must be included in the model.
This interaction results from the electron-lattice coupling and
is a generalization of the JT interaction which
was found to play an important role in the parent LaMnO$_3$ 
compound.\cite{Mil96,Ahn98} In CO compounds at quarter-filling
the effect of the OO 
coupling will be dominated by interactions between  further distant Mn
neighbors. We have shown that such  
interactions can produce the CE structure in both {\it spin and 
orbital} sectors. In the
case of the three-site effective interaction ($\sim \kappa'$) 
the zigzag chains are found to be
stable not only for modest inter-site repulsion ($V\simeq 0.5t$) but even in 
the limit of $V=0$ and provided $\kappa$ is small, where the CO is induced 
entirely by the three-site
orbital coupling (\ref{eq:3site}). This effective next-nearest-neighbor OO
coupling was derived taking into account the cooperative
nature of both Mn$^{3+}$--O and Mn$^{3+}$--Mn$^{4+}$ bonds length 
deformation. The free relaxation of the latter length  leads
in presence of CO to orbital alternation in planar directions, 
and the pattern of directed orbitals typical for the orbital CE
structure appears.
Consequently, with the help of the orbital 
topology,\cite{Sol99,BKK99} this gives rise to the 
CE AF spin ordering for realistic values for $J_\textrm{AF}$.

An important aspect of the second neighbor JT-interaction (iii) is its
3-site structure which involves a coupling to the charge at the 
Mn$^{4+}$ ion. This implies that the evolution of the CE orbital 
correlations with temperature are linked to the evolution of CO, 
while the magnetic CE correlations develop at lower temperature
similar to  Pr$_{1/2}$Ca$_{1/2}$MnO$_3$\cite{Kaj01,zim99}.
This is different from (ii), where first CO develops, then OO and
finally at lowest temperature AF spin order. 

As discussed in Sec.~\ref{sec:parameters} we did not find 
the CE spin and/or
orbital phase considering a pure electronic model, i.e., neglecting 
the electron-lattice coupling. Although, the topological effect combined
with the DE mechanism leads to the preference of the CE over C phase
when all $e_g$ (bridge and corner) orbitals are degenerate,\cite{BKK99}
the situation 
is more complex when such degeneracy is lifted by JT lattice distortions.
As described in the Appendix, one finds band insulators 
in both cases
while the stability of the CE versus the C phase strongly depends on 
the crystal-field splittings ($E_z$, $\eta E_z\sim\lambda$). 

An important issue is the degree of charge ordering in the CE phase. 
The CE spin-orbital structures obtained in our simulations appear
in the regime of relatively strong CO, as infered from our
calculated charge correlation functions. Our
calculations also suggest that the second neighbor JT coupling even for small
or vanishing Coulomb repulsion can lead to a pronounced charge modulation.
These results are consistent with the simple physical picture,
that DE perpendicular to the FM chains must be strongly
suppressed to allow AF superexchange to trigger the CE structure, which may
not be the case for weak CO.
On the other hand, a recent study by Mahadevan, Terakura and Sarma\cite{Mah01}
based on the band structure approach arrived at the conclusion that the
charge difference between the two Mn species in the CE phase is 
basically negligible, which would imply the absence of a significant breathing 
distortion of Mn$^{4+}$ octahedra.
Further experimental information on the degree of lattice distortion
in the CE phase will certainly help to test the proposed JT based 
mechanism for the stability of the CE phase.

The charge modulations determined in our calculations may be influenced by
numerical limitations which overestimate quantum fluctuations:
(i) spins $S=1/2$ are used to describe $t_{2g}$ electrons; 
(ii) the calculations are made using a 2D cluster;
(iii) the oxygen ions are not present explicitly in the cluster and thus 
only \textit{dynamical} interactions are included ($\kappa,\kappa'$),
which are mediated by oxygen displacements, while
possible \textit{static} lattice effects\cite{Bal00} are neglected. 
Certainly each of these approximations weakens 
the charge ordering and the spin-orbital order. 
It is, however, not straightforward to decide whether 
the region of the CE-phase in the phase diagram may grow or shrink, if 
quantum fluctuations are suppressed.

Finally we note that further neighbor JT
interactions can also play an important role in the charge stripe 
formation at other commensurate carrier concentrations. For $x<1/2$ ($x>1/2$)
the dominant coupling arises from the nearest- (next nearest-) neighbor 
OO interactions, respectively, with a pronounced electron-hole asymmetry
present. In the latter case with small inter-site Coulomb repulsion
($V$) the bi-stripe state\cite{Mor98} 
would be preferred while larger values of $V$ would favor 
a ``Wigner crystal'' charge arrangement.\cite{Rad99}

%%%%%%%%%%%%%%%%%%%%%%%%%%%%%%%%%%%%%%%%%%%%%%%%%%%%%%%%%%%%%%%%%%%%%%%%%%%%
%%
%%                           ACKNOWLEDGMENTS
%%
%%%%%%%%%%%%%%%%%%%%%%%%%%%%%%%%%%%%%%%%%%%%%%%%%%%%%%%%%%%%%%%%%%%%%%%%%%%%
\begin{acknowledgments}

We would like to thank G. Khaliullin, M. Mayr and A. M. Ole\'s for 
stimulating discussions and valuable comments.
J. B. acknowledges the support of the MPI f\"ur Festk\"orperforschung,
Stuttgart. The financial support by the Polish State Committee 
of Scientific Research (KBN) of Poland, Project No. 5~P03B~055~20
is also acknowledged (J. B.).

\end{acknowledgments}
%%%%%%%%%%%%%%%%%%%%%%%%%%%%%%%%%%%%%%%%%%%%%%%%%%%%%%%%%%%%%%%%%%%%%%%%%%%%
%%
%%                           APPENDICES
%%
%%%%%%%%%%%%%%%%%%%%%%%%%%%%%%%%%%%%%%%%%%%%%%%%%%%%%%%%%%%%%%%%%%%%%%%%%%%%
\appendix*

%%%%%%%%%%%%%%%%%%%%%%%%%%%%%%%%%%%%%%%%%%%%%%%%%%%%%%%%%%%%%%%%%%%%%%%%%%%%
\section{\label{sec:1D}1D band model}
%%%%%%%%%%%%%%%%%%%%%%%%%%%%%%%%%%%%%%%%%%%%%%%%%%%%%%%%%%%%%%%%%%%%%%%%%%%%

A quite appealing explanation for the stability of the CE as compared to the
C phase was suggested in Refs.~\onlinecite{Sol99,BKK99} by considering a 
1D model, where in the CE phase a gap opens as a consequence of
a topological sign in the hopping matrix-elements. 
However, the intra-site Coulomb repulsion $U$ (leading to some
redistribution of the $e_g$ charge) can destabilize the CE 
phase.\cite{She01,Cuo02} 
Here we investigate how 
these arguments are modified when orbital degeneracy is lifted due to 
JT distortions. We consider the Hamiltonian: 
\begin{equation}
H^\textrm{1D} = H_t + E_{z(b)}\sum_{i\in corner}T^z_i 
+ \frac{1}{2}\eta E_{z(b)}\sum_{i\in bridge}n_i ,
\label{eq:H^1D}
\end{equation}
which contains the kinetic energy $H_t$ of an electron hopping along a FM 
chain\cite{BKK99} 
with possible $e_g$ level shift at the bridge sites ($\propto\eta E_{z(b)}$) 
and level splitting at corner sites ($\propto E_{z(b)}$) considered in two 
different orbital basis $\{|x\rangle,|z\rangle\}$ and 
$\{|a\rangle=\frac{1}{\sqrt{2}}(|x\rangle+|z\rangle),
|b\rangle=\frac{1}{\sqrt{2}}(|x\rangle-|z\rangle)\}$. 
We assume that the energies of the bridge 
$3x^2-r^2$/$3y^2-r^2$ orbitals are located between the split energy levels
at the corners 
assuming $|\eta|\leq 1$. Both kinds of level shifts can result 
from the lattice distortions. At the bridge site only the 
orbital $3x^2-r^2$ ($3y^2-r^2$) is considered while the orthogonal one 
$y^2-z^2$ ($x^2-z^2$), respectively, is decoupled in  chain direction.
In  momentum space this leads to a $3\times 3$ matrix problem:
\begin{equation}
\label{eq:1D_bands}
H^{1D} = \sum_{k}
\left(\begin{array}{c}
d^{\dagger}_{B,k}\\ d^{\dagger}_{x,k}\\ d^{\dagger}_{z,k}\end{array}
\right)^T
\left[\begin{array}{ccc}
\eta E_z & W'_k & W_k\\
W'_k & E_z & 0\\
W_k & 0 & -E_z
\end{array}
\right]
\left(\begin{array}{c}
d_{B,k}\\ d_{x,k}\\ d_{z,k}\end{array}
\right) ,
\end{equation}
where $W_k=-2t\cos k$ ($2t\cos k$) and 
$W'_k=-2\sqrt{3}t\sin k$ ($2\sqrt{3}t\cos k$) are the hopping elements 
for the CE (C) phase, respectively, in the 
$\{|x\rangle,|z\rangle\}$ orbital basis. 

%%%%%%%%%%%%%%%%%%%%%%%%%%%%%%%% FIG. 15 %%%%%%%%%%%%%%%%%%%%%%%%%%%%%%%%%%%
\begin{figure}
\includegraphics[width=5cm]{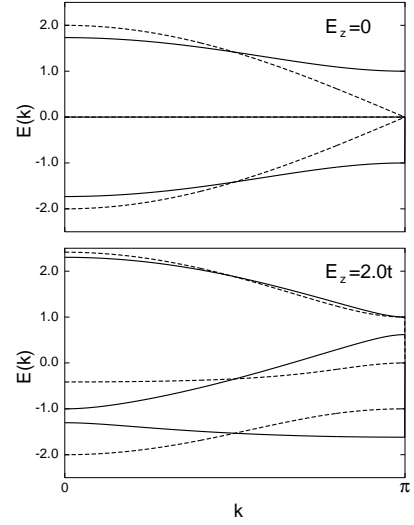}
\caption{\label{fig:bands}
Electron dispersion for the zigzag chain of the CE phase (solid
lines) and for the linear chain of the C phase (dashed lines) calculated for
different crystal-field splittings in the $\{|x\rangle,|z\rangle\}$ 
orbital basis at corner sites ($E_z$). The 
orbitals at the bridge sites are degenerate (assuming $\eta=0$).}
\end{figure}
%%%%%%%%%%%%%%%%%%%%%%%%%%%%%%%%%%%%%%%%%%%%%%%%%%%%%%%%%%%%%%%%%%%%%%%%%%%%
%%%%%%%%%%%%%%%%%%%%%%%%%%%%%%%% FIG. 16 %%%%%%%%%%%%%%%%%%%%%%%%%%%%%%%%%%%
\begin{figure}
\includegraphics[width=5cm]{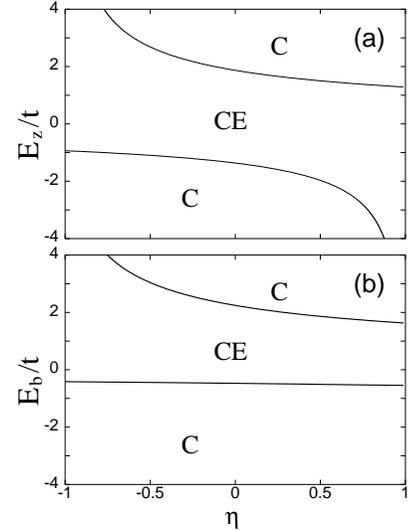}
\caption{\label{fig:1D_phase}
Phase diagram of the chain model illustrating the stability
of the CE against the C phase as function of the  crystal-field splittings 
defined with respect to the orbital basis
(a) $\{|x\rangle,|z\rangle\}$ and (b) 
$\{\frac{1}{\sqrt{2}}(|x\rangle\pm|z\rangle)\}$, respectively.}
\end{figure}
%%%%%%%%%%%%%%%%%%%%%%%%%%%%%%%%%%%%%%%%%%%%%%%%%%%%%%%%%%%%%%%%%%%%%%%%%%%%

Calculating the band energies (see Fig.~\ref{fig:bands}) 
one can easily determine the character of the stable 
phase. For the degenerate band model ($E_z=0$) the C phase is 
metallic while in the CE phase a gap opens due to the \textit{topological 
sign}.\cite{Sol99,BKK99} 
The opening of the gap stabilizes the CE phase.
When the  degeneracy  between $|x\rangle$ and $|z\rangle$ states 
at the corner sites ($E_z\neq 0$) is lifted, the C phase becomes insulating 
as well and for sufficiently large $e_g$ level splittings it can have 
a \textit{lower} energy than the CE phase when the chain is doped with 
one electron per two Mn sites [see Fig.~\ref{fig:bands}(b)]. 
The region of stability of the CE phase increases
for positive (negative) $E_z$ when $\eta\to -1$($1$), 
respectively [see Fig.~\ref{fig:1D_phase}(a)]. Generally, the CE phase is 
stabilized with respect to C, when the  occupied bridge state
approaches the  lower energy level of the corner states.
In a real half-doped crystal where in the CE phase Mn$-$O bond lengths 
shrink (expand) in the $c$ direction [$(a,b)$ plane], respectively,
one would expect an energetic preference, due to the JT effect,
for orbitals with stronger $x^2-y^2$ than $3z^2-r^2$ character. 
Thus, in the 1D model the relevant part of the phase diagram 
(Fig.~\ref{fig:1D_phase}) is the region with $E_z<0$ and $\eta E_z<0$, 
i.e., where the CE structure dominates. 

To relate our mean-field considerations to the model from Sec.~\ref{sec:nnjt}
the stability of the CE phase versus C was also considered with level
splitting at the corner sites defined with respect to the 
orbital basis $\{\frac{1}{\sqrt{2}}(|x\rangle\pm|z\rangle)\}$.
Such a splitting, directly related to the nearest-neighbor JT interaction
(\ref{eq:H_OO}) when $E_b>0$, 
is operative in the C phase but vanishes by symmetry
in the CE orbital phase. As shown in Fig.~\ref{fig:1D_phase}(b) the CE phase
is stable for small splittings ($0\alt E_b\alt 2t$) 
and in the  $E_b>0$, $\eta\to -1$ region. The latter is the relevant parameter
range in the case of nearest neighbor JT interactions, and hence the model 
suggests the CE phase as stable phase. This, however, is in conflict with
our numerical study of the 2D model, where only the C phase was found to be 
stable. As in 2D the nearest neighbor JT interaction stabilizes the C
structure, whereas this interaction is frustrated in the CE phase. 
This effect is not included in the 1D model.

%%%%%%%%%%%%%%%%%%%%%%%%%%%%%%%%%%%%%%%%%%%%%%%%%%%%%%%%%%%%%%%%%%%%%%%%%%%%
%%
%%                           REFERENCES
%%
%%%%%%%%%%%%%%%%%%%%%%%%%%%%%%%%%%%%%%%%%%%%%%%%%%%%%%%%%%%%%%%%%%%%%%%%%%%%


\begin{thebibliography}{99}
\bibitem{noncentral} To account for local non-central OO interactions
                     one should use full sets of coordinates for each 
                     ion in Mn$^{4+}$ octahedra and include nonlinear 
                     lattice effects.
\bibitem{dimension} We note that $K_1=\tilde{K_1} b^2$ and 
$\lambda=\tilde{\lambda} b$; hence in conventional notation these values 
correspond to $\tilde{K_1}\approx 13$ eV/{\AA}$^2$ and 
$\tilde{\lambda}\approx 1.5$ eV/\AA.
\end{thebibliography}
\end{document}